\documentclass[aps,prc,twocolumn,amsmath,amssymb,superscriptaddress,showpacs,showkeys]{revtex4}
\usepackage{dcolumn}
\usepackage{graphicx,color,bm}
\usepackage{textcomp}

\begin {document}
  \newcommand {\nc} {\newcommand}
  \nc {\beq} {\begin{eqnarray}}
  \nc {\eeq} {\nonumber \end{eqnarray}}
  \nc {\eeqn}[1] {\label {#1} \end{eqnarray}}
  \nc {\eol} {\nonumber \\}
  \nc {\eoln}[1] {\label {#1} \\}
  \nc {\ve} [1] {\mbox{\boldmath $#1$}}
  \nc {\ves} [1] {\mbox{\boldmath ${\scriptstyle #1}$}}
  \nc {\mrm} [1] {\mathrm{#1}}
  \nc {\half} {\mbox{$\frac{1}{2}$}}
  \nc {\thal} {\mbox{$\frac{3}{2}$}}
  \nc {\fial} {\mbox{$\frac{5}{2}$}}
  \nc {\la} {\mbox{$\langle$}}
  \nc {\ra} {\mbox{$\rangle$}}
  \nc {\etal} {\emph{et al.}}
  \nc {\eq} [1] {(\ref{#1})}
  \nc {\Eq} [1] {Eq.~(\ref{#1})}
  \nc {\Ref} [1] {Ref.~\cite{#1}}
  \nc {\Refc} [2] {Refs.~\cite[#1]{#2}}
  \nc {\Sec} [1] {Sec.~\ref{#1}}
  \nc {\chap} [1] {Chapter~\ref{#1}}
  \nc {\anx} [1] {Appendix~\ref{#1}}
  \nc {\tbl} [1] {Table~\ref{#1}}
  \nc {\fig} [1] {Fig.~\ref{#1}}
  \nc {\ex} [1] {$^{#1}$}
  \nc {\Sch} {Schr\"odinger }
  \nc {\flim} [2] {\mathop{\longrightarrow}\limits_{{#1}\rightarrow{#2}}}
  \nc {\textdegr}{$^{\circ}$}
  \nc {\inred} [1]{\textcolor{red}{#1}}
  \nc {\inblue} [1]{\textcolor{blue}{#1}}
\nc {\IR} [1] {\textcolor{red}{#1}}
\nc {\IB} [1] {\textcolor{blue}{#1}}
  \nc {\efbcom}[1]{\textcolor{blue}{\bfseries\itshape #1}}
  \nc {\GK} {\ensuremath{\mathrm{\,GK}}}
  \nc {\Msun} {\ensuremath{\,M_{\odot}}}
\title{Investigation of the triple-alpha reaction in a full three-body approach}
\author{N.~B.~Nguyen}
\email{nguyenn@nscl.msu.edu}
\affiliation{National Superconducting Cyclotron Laboratory
and Department of Physics and Astronomy,
Michigan State University, East Lansing, Michigan 48824, USA}
\author{F.~M.~Nunes}
\email{nunes@nscl.msu.edu}
\affiliation{National Superconducting Cyclotron Laboratory
and Department of Physics and Astronomy,
Michigan State University, East Lansing, Michigan 48824, USA}
\author{I.~J.~Thompson}
\affiliation{Lawrence Livermore National Laboratory, L-414,
 Livermore, California 94551, USA}

\date{\today}
\begin{abstract}
{\bf Background:} The triple-alpha reaction is the key to our understanding about the nucleosynthesis and the observed abundance of $^{12}$C in stars. The theory of this process is well established at high temperatures but rather ambiguous in the low temperature regime where measurements are impossible. {\bf Purpose:} Develop a new three-body method, which tackles properly the scattering boundary condition for three charged particles and takes into account both the resonant and the non-resonant reaction mechanisms on the same footing, to compute the triple-alpha reaction rate at low temperatures. {\bf Methods:}  We combine the R-matrix expansion, the R-matrix propagation method, and the screening technique in the hyperspherical harmonics basis. {\bf Results: } Both the $2^+_1$ bound state and the $0^+_2$ resonant state in $^{12}$C are well reproduced. We also study the cluster structure of these states. We calculate the triple-alpha reaction rate for $\mathrm{T}=0.01-0.1$ GK. {\bf Conclusions:} We obtain the same rate as NACRE for temperatures above $0.07$ GK, but the new rate is largely enhanced at lower temperatures ($\approx 10^{12}$ at $0.02$ GK). The differences are caused by the direct capture contribution to the reaction when three alpha particles can not reach the resonant energies. 
\end{abstract}
\pacs{24.50.+g, 25.10.+s, 25.40.Lw, 26.20.Fj}
\keywords{triple-alpha reaction rate, $^{12}$C, hyperspherical harmonics, R-matrix, three charged-particle fusion}
\maketitle

\section{Introduction}
The formation of $^{12}$C through the triple-alpha reaction was first suggested by Salpeter to explain the synthesis of heavy nuclei since stable isotopes of mass number 5 and 8 can not be found in nature \cite{salpeter}. However, this reaction rate was too small to account for the observed abundance of $^{12}$C in stars. Later on, Fred Hoyle predicted the existence of a $0^+$ resonance near the three-alpha threshold which significantly increased the reaction rate \cite{hoyle54}. This resonant state and its properties were confirmed by experimentalists \cite{dunbar53, cook57} shortly thereafter. Then, the two-step (sequential) mechanism of the triple-alpha capture,
\begin{eqnarray}
\alpha+\alpha \rightarrow \mathrm{^8 Be}(0^+_1)\\
\mathrm{^8Be}(0^+_1)+\alpha \rightarrow \mathrm{^{12}C}(0^+_2)\, ,
\end{eqnarray} 
was accepted to explain the mystery of nucleosynthesis and the abundance of $^{12}$C in helium burning stars. \\
\indent The first evaluation of the sequential (resonant) triple-alpha reaction rate was credited to Caughlan and Fowler \cite{caughlan88} in which only the Hoyle resonance was considered and a Breit-Wigner shape for the resonant cross sections of $\alpha-\alpha$ and $\alpha-^{8}$Be was used. Ten years later, this reaction rate  was improved by Angulo ~\cite{angulo99} when taking into account the contribution of the $2_2^+$ resonance. This rate is included in NACRE: \underline{N}uclear \underline{A}strophysics \underline{C}ompilation of \underline{RE}action Rates. At low temperatures, three alpha particles do not have access to the intermediate resonances, so one expects the three-alpha direct capture to be dominant. The non-resonant triple-alpha rate tabulated in NACRE is obtained by a simple extrapolation of the sequential model to low energies. Since at low temperatures the measurements are impossible, a well founded theory for the three-alpha direct (non-resonant) capture is crucial.\\
\indent Ogata {\it et al.} \cite{ogata09} was the first group who attempted to tackle this problem without using an ambiguous extrapolation. In their studies, the continuum discretized coupled channel (CDCC) method \cite{cdcc} was employed to solve a three-body scattering equation, offering an opportunity for the inclusion of both resonant and non-resonant mechanisms in the triple-alpha reaction. However, the CDCC results \cite{ogata09} drastically differ from NACRE at low temperatures (20 orders of magnitude at $T=0.02$ GK) which significantly affects astrophysical studies and produces results that are incompatible with observation \cite{dotter09,suda11,white-dwarfs,neutron-stars,peng-xrb}. Applying the CDCC method to the triple-alpha problem can be challenging because the CDCC wavefunction which is usually truncated can not reproduce the correct scattering behavior of three charged particles in the asymptotic region \cite{vasilevsky01}.\\
\indent Another attempt using the hyperspherical adiabatic expansion method to solve the triple-alpha problem was made by the Madrid-Aarhus collaboration \cite{alvarez07,diego10}. However, they were unable to perform the numerical calculation at temperatures below $T\approx 0.1$ GK. Therefore, in order to estimate the non-resonant triple-alpha rate at low temperatures, they extrapolated a three-body Breit-Wigner cross section for the three-alpha capture to low energies \cite{garrido11} in a similar manner as done in the sequential process \cite{angulo99}. The reaction rate from this work (named as BW(3B)) shows an increase of 7 orders of magnitude at $T=0.02$ GK compared to NACRE. There are large discrepancies in the results of \cite{angulo99}, \cite{ogata09} and \cite{garrido11}, demonstrating uncertainties and ambiguities in our understanding of the low temperature triple-alpha reaction. The aim of our work is to resolve this problem.\\
\indent At low temperature, the triple-alpha reaction proceeds through the $0^+$ continuum states to the $2^+_1$ bound state of $^{12}$C. In order to construct a reliable non-resonant reaction rate, we need a good description of both $^{12}$C bound and continuum states. Recent  microscopic theories such as the no-core shell model \cite{maris09} and the Green's function Monte Carlo method \cite{pieper05} failed to reproduce the Hoyle state of $^{12}$C by solving a true 12-body problem. Even though the fermionic molecular dynamics method produces both the $2^+_1$ bound state as well as the $0^+_2$ resonance with significant triple-alpha configurations \cite{chernykh07}, a microscopic description for the non-resonant continuum states is currently not available. Given the dominant triple-alpha structure in both $^{12}$C($2^+_1$) and $^{12}$C($0^+_2$), it is reasonable to construct this as a three-body problem as done in \cite{ogata09, diego10, alvarez07}.\\
\indent Theories for the three charged particle problem have made significant progress over the last few decades. An accurate solution for a three-body bound state system can now be obtained (e.g \cite{face}). However, there are still difficulties remaining for a three charged particle scattering problem because no standard boundary condition for this system exists when considering the long-range effects of the Coulomb interaction. Some efforts have been made to solve this scattering problem but the results are limited to cases where only one pair of particles have charge \cite{kievsky97, deltuva05}. Our triple-alpha problem is more difficult since all three particles are charged. In this work we employ the hyperspherical harmonics (HH) method in combination with the R-matrix expansion and the R-matrix propagation technique to provide a good description of the $\alpha+\alpha+\alpha$ system at low relative energies with the correct Coulomb asymptotic behavior, thereby enabling us to compute the reaction rate at very low temperatures without extrapolation. Our new three-body method is named here as HHR. This article is an extended version of the letter in \cite{nguyen12}, in which more details on the theoretical methods are presented and a more comprehensive analysis of the results is performed.\\
\indent We have organized this article in the following way: detailed theory on the method is presented in Sec.~II; the main results are shown in Sec.~III. In Sec.~IV we discuss in depth the reaction dynamics and the conclusions are drawn in Sec.~V.



\section{Theory}
\subsection{Hyperspherical formalism \label{HHform}}
The hyperspherical harmonics (HH) method originated in atomic and molecular physics \cite{gronwall37, lin95}. Later, it was extended to few-body systems in nuclear physics by Delves \cite{delves58, delves62}. The theory of the HH method was well developed for Borromean systems \cite{zhukov93, tarutina04, descouvemont03}. Borromean nuclei are defined as systems of three particles which are loosely bound and have no bound states in any of the two-body subsystems. $^{11}$Li and $^{6}$He which have two neutrons weakly coupled to the core, are typical examples of Borromean nuclei. In this work, we formulate our problem as a Borromean system of three alpha particles.\\
\indent Let us consider a system of three nuclei with masses $m_i$. For each Jacobi set $i$, we define the scaled coordinates ${\bf x}_i=\sqrt{\frac{A_jA_k}{A_j+A_k}}\:{\bf r}_{i}$ and ${\bf y}_i =\sqrt{\frac{A_i(A_j+A_k)}{A_i+A_j+A_k}}\:{\bf R}_{i}$. Here, ${\bf r}_{i}$ is the relative radius vector from particle $j$ to particle $k$ and ${\bf R}_{i}$ is the radius vector from particle $i$ to the center of mass of the two-body subsystem ($jk$). The ratios $A_i=m_i/m$ are dimensionless and the scaled mass $m$ is usually taken as the nucleon mass. The hyperspherical coordinates are then defined as functions of $x_i$ and $y_i$
\begin{eqnarray}
\rho^2 &=&x_i^2+y_i^2 \, ,\nonumber\\
\theta_i &=&\mathrm{arctan}\:\frac{x_i}{y_i} \, .
\label{hyperco-eq}
\end{eqnarray}
The hyper-radius $\rho$ is invariant under translations, rotations, and permutations while the hyper-angle $\theta_i$ depends on the selected Jacobi set. For the triple-alpha problem in which three particles are identical, all the Jacobi coordinate sets are equivalent, so we choose to work with Jacobi set $i=3$ and drop the index $i$ in our equations from now on for convenience.\\
\indent Because our problem involves only spin-zero particles, all the equations in HH coordinates are written for this simpler form. The general case of non-zero spin particles is presented with details in \cite{nunes96,book}. The hyperspherical expansion separates the radial and angular dependence of the three-body wavefunction
\begin{equation}
\Psi^{LM}=\rho^{-\frac{5}{2}}\sum_{Kl_xl_y}\chi_{Kl_xl_y}(\rho)\:\varphi_K^{l_xl_y}(\theta)\:[Y_{l_x}\otimes Y_{l_y}]_{LM} \, .\label{hhexpan-eq}
\end{equation}
In Eq.~(\ref{hhexpan-eq}) $\chi_{Kl_xl_y}(\rho)$ are the hyper-radial functions, solutions of the coupled channels equations (\ref{hh-eq}). The hyper-angular functions $\varphi_K^{l_xl_y}(\theta)$ are the eigensolutions of the hyper-angle equation and are defined by Jacobi polynomials \cite{nunes96}. This expansion introduces a new quantum number, the hyper-momentum $K$, an extended concept of angular momentum for a three-body system. $l_x$ and $l_y$ are the orbital angular momenta corresponding to the Jacobi coordinates $x$ and $y$. \\
\indent Using the HH expansion above, we arrive at a set of coupled channels equations in the hyper-radius coordinate:
\begin{equation}
\left(\frac{\hbar^2}{2m}\left[\frac{\mathrm{d}^2}{\mathrm{d}\rho^2}{-}\frac{\Delta(\Delta{+}1)}{\rho^2}\right]{+}E\!\right)\:\chi_{\gamma}(\rho)=
\sum_{\gamma'}V_{\gamma \gamma'}(\rho)\:\chi_{\gamma'}(\rho),
\label{hh-eq}
\end{equation}
where $\Delta=K+3/2$ and $\gamma=\{K, l_x, l_y\}$. The coupling potentials $V_{\gamma \gamma'}(\rho)$ are defined as the sum of three pairwise interactions $V_{jk}(\rho,\theta)$ plus a three-body force $V_{3b}(\rho)$, integrated over all variables but $\rho$:
\begin{equation}
V_{\gamma \gamma'}(\rho)=
\langle \Omega_{\gamma}(\theta,\hat{\bf x},\hat{\bf y})|\sum_{k>j=1}^3\!\! V_{jk}+V_{3b}|\Omega_{\gamma'}(\theta,\hat{\bf x},\hat{\bf y})\rangle
\; ,\label{coupl-eq}
\end{equation}
where $\Omega_{\gamma}(\theta,\hat{\bf x},\hat{\bf y})=\varphi_K^{l_xl_y}(\theta)\:[Y_{l_x}\otimes Y_{l_y}]$ are the hyperharmonic basis functions containing all the angular dependence in Eq.~(\ref{hhexpan-eq}). In the HH representation, both the diagonal and the off-diagonal couplings decay slowly as $Z^{\mathrm{eff}}_{\gamma\gamma'}/\rho$ for a system of three charged particles \cite{vasilevsky01}, demonstrating the difficulty of our probem in the asymptotic region. $Z^{\mathrm{eff}}_{\gamma\gamma'}$ is constant for any given channel $\gamma$ and $\gamma'$ and has a non-trivial expression:
\begin{eqnarray}
&&Z^{\mathrm{eff}}_{\gamma\gamma'}=Z^2e^2\sqrt{\frac{A}{2}}\left[\:\left\langle\gamma,3\left|\frac{1}{\mathrm{sin}\theta_3}\right|\gamma',3\right\rangle\right.\nonumber\\
&&+\sum_{\alpha\alpha'}\Re_{\gamma\alpha}^{32}\Re_{\alpha'\gamma'}^{23}\left\langle\alpha,2\left|\frac{1}{\mathrm{sin}\theta_2}\right|\alpha',2\right\rangle\nonumber\\
&&+\left.\sum_{\beta\beta'}\Re_{\gamma\beta}^{31}\Re_{\beta'\gamma'}^{13}\left\langle\beta,1\left|\frac{1}{\mathrm{sin}\theta_1}\right|\beta',1\right\rangle\right]\, ,
\end{eqnarray}
where $Z$ and $A$ are the charge number and the atomic number of an alpha particle; $\Re_{\beta\gamma}^{ij}$ is the Raynal-Revai coefficient \cite{raynal}; $i=1$, 2, 3 represent three Jacobi coordinates ($\bf{x}_i,\bf{y}_i$) and $|\gamma,i\rangle$ is the hyperharmonic basis function $\Omega_{\gamma}(\theta_i,\hat{{\bf x}_i},\hat{{\bf y}_i})$ in Jacobi $i$. Each matrix element is integrated over all angular variables.\\
\indent For neutral Borromean systems, the three-body bound state wavefunction decays exponentially at large distance \cite{merkuriev76}. When introducing Coulomb interactions the problem becomes more difficult because a simple analytic expression for the asymptotic bound state wavefunction of three charged particles does not exist. However, for a well-bound system, imposing the  boundary condition $\chi_{\gamma}(\rho \rightarrow\infty)\rightarrow 0$ is sufficient for the numerical calculation \cite{nunes96}.\\
\indent Solving Eq.~(\ref{hh-eq}) for positive energies $E$, to obtain continuum states for a system of three charged particles, is numerically challenging since it requires an exact boundary condition. Fortunately, there are techniques (which will be discussed in detail later) that enable us to neglect the off-diagonal potential couplings at large distances. This results in analytic solutions for the asymptotic version of Eq.~(\ref{hh-eq}) corresponding to the regular $F_{\Delta}(\eta_{\gamma},\kappa\rho)$ and irregular $G_{\Delta}(\eta_{\gamma},\kappa\rho)$ three-body Coulomb functions. We can thus write down the asymptotic three-body continuum wavefunction for a charged system in hyper-radius:
\begin{equation}
\chi_{\gamma\gamma'}^{\mathrm{HHR}}(\rho)\stackrel{{\rho\to\infty}}{\longrightarrow} H^{-}_{\Delta}(\eta_{\gamma},\kappa\rho)\delta_{\gamma\gamma'}- H^{+}_{\Delta}(\eta_{\gamma},\kappa\rho)S_{\gamma\gamma'}\, .\label{coulwf-eq}
\end{equation}  
$H^{\pm}=G\pm iF$ are the Coulomb functions describing the outgoing and incoming spherical waves and $S_{\gamma\gamma'}$ is the scattering matrix \cite{book}. Due to the nature of the scattering boundary condition, each partial wave $\chi_{\gamma\gamma'}$ requires two subscripts $\gamma$ and $\gamma'$ to describe the outgoing and incoming channels respectively. The hyper-radial momentum $\kappa$ is given as a function of the three-body kinetic energy $\kappa=\sqrt{2mE/\hbar^2}$, where $\hbar$ is the Planck constant. The variable $\eta_{\gamma}$ is an equivalent of the Sommerfeld parameter in two-body systems but is more complex and hyper-momentum $K$ dependent. $\eta_{\gamma}$ is obtained from the parameter $Z^{\mathrm{eff}}_{\gamma\gamma}$ through a relationship $\eta_{\gamma}=mZ^{\mathrm{eff}}_{\gamma\gamma}/\hbar^2\kappa$.
\subsection{R-matrix expansion}
Wigner and Eisenbud \cite{wigner47} were the first to introduce the R-matrix method into nuclear physics to study resonances in nuclear scattering. Later on, this method was developed and extended in detail by Lane and Thomas \cite{lane58}. The general idea of this method is to use an orthonormal basis expansion inside an R-matrix box. An R matrix is then constructed to match to the asymptotic wavefunction outside the box \cite{book}. Even though the R-matrix method was originally developed for a two-body scattering problem, it has been generalized for a three-body system in HH coordinates by I.J. Thompson {\it el al.} \cite{hh-R-matrix}. Typically the R-matrix expansion provides a method that is numerically more stable than direct integration methods for solving coupled channels equations with strong repulsive couplings \cite{book}.\\
\indent By imposing a constant logarithmic derivative $\beta=\mathrm{dln}\phi_{\gamma}^p(\rho)/\mathrm{d}\rho$ at a fixed radius $\rho_m$, the so called R-matrix radius, the eigensolutions $\phi_{\gamma}^p(\rho)$ of Eq.~(\ref{hh-eq}) will form an orthonormal basis over a finite range $[0,\rho_m]$. The scattering wavefunctions at energy E are expanded in terms of $\phi_{\gamma}^p(\rho)$ at eigenenergies $e_p$:
\begin{equation}
\chi_{\gamma\gamma'}=\sum_p A_{\gamma\gamma'}^p\:\phi_{\gamma}^p\, .
\label{wfsrmtrx1-eq}
\end{equation}
In our calculation, the logarithmic derivative $\beta$ is chosen to be zero. This option may lead to the discontinuity of the derivative of the wavefunction at the R-matrix boundary when using a truncated basis as discussed in \cite{descouvemont10}. However, a sufficiently large basis is employed in our calculation to ensure a smooth behavior of the wavefunction at all energies. In addition, the zero logarithmic derivative highly reduces the complexity of the R-matrix propagation method (Sec.~\ref{r-propa}). All the equations presented in this paper consider $\beta=0$ while more general formulae can be found in \cite{book}. The R matrix at energy $E$ relates the wavefunction $\chi_{\gamma\gamma'}(\rho)$ and its derivative: 
\begin{equation}
\chi_{\gamma\gamma'}(\rho)=\sum_{\gamma''}\rho \:R_{\gamma\gamma''}(E)\:\chi_{\gamma''\gamma'}'(\rho)\, .
\label{wfsrmtrx2-eq}
\end{equation}
The constant logarithmic derivative enables us to calculate the R matrix $R_{\gamma\gamma''}(E)$ from the known eigenfunctions $\phi_{\gamma}^p$ at radius $\rho_m$:
\begin{equation}
R_{\gamma\gamma''}(E)=\frac{\hbar^2}{2m\rho_m}\:\sum_{p=1}^P\frac{\phi_{\gamma}^p(\rho_m)\:\phi_{\gamma''}^p(\rho_m)}{e_p -E}\, ,
\label{rmtrx-eq}
\end{equation}
where $P$ is the number of poles used in our calculation. If we substitute Eq.~(\ref{wfsrmtrx1-eq}) and Eq.~(\ref{rmtrx-eq}) into Eq.~(\ref{wfsrmtrx2-eq}), the only missing quantities to calculate the scattering wavefunction are the expansion coefficients $A_{\gamma\gamma'}^p$. As discussed in Sec.~\ref{HHform}, Eq.~(\ref{coulwf-eq}) is the asymptotic form of a continuum wavefunction for three charged particles when the off-diagonal couplings can be neglected. For sufficiently large $\rho_m$ the wavefunction will be in the asymptotic regime. Inserting Eq.~(\ref{wfsrmtrx1-eq}) and Eq.~(\ref{rmtrx-eq}) in Eq.~(\ref{wfsrmtrx2-eq}) and combining with Eq.~(\ref{coulwf-eq}) we obtain the unknown coefficients $A_{\gamma\gamma'}^p$:
\begin{equation}
A_{\gamma\gamma'}^p=\frac{\hbar^2}{2m}\:\frac{1}{e_p -E}\:\sum_{\gamma''}\phi_{\gamma''}^p[\delta_{\gamma''\gamma'}{H'}_{\gamma''}^{-}-S_{\gamma''\gamma'}{H'}_{\gamma''}^{+}]\, .
\label{coefexp-eq}
\end{equation}
All functions in Eq.~(\ref{coefexp-eq}) are evaluated at the R-matrix boundary $\rho_m$. The scattering matrix $S_{\gamma''\gamma'}$ and the R matrix are directly related \cite{book}.
\subsection{R-matrix propagation \label{r-propa}}
The triple-alpha system is driven by strong, long-range pairwise Coulomb interactions. Therefore, the R-matrix radius $\rho_m$ has to be very large to fully capture the physics of the problem. Solving the coupled channels equations in a large R-matrix box will lead to numerical instability. In order to overcome this difficulty, we employ the R-matrix propagation technique originally developed by Light and Walker for the atom-molecule scattering problem \cite{light76, light78}. First, the set of coupled channels equations (\ref{hh-eq}) is solved in a small R-matrix box with a radius of $\rho_m$. Then the R matrix at $\rho_m$ is propagated to $\rho_a\gg \rho_m$ where matching to known Coulomb functions can be performed. This propagation method is well known for its fast convergence and high numerical stability.\\
\indent We rewrite Eq.~(\ref{hh-eq}) as
\begin{equation}
\frac{\mathrm{d}^2\chi_{\gamma\gamma'}(\rho)}{\mathrm{d}\rho^2}=\sum_{\gamma''}\tilde{V}_{\gamma\gamma''}\:\chi_{\gamma''\gamma'}(\rho)\,.
\label{hhpropa-eq}
\end{equation}
The interval from the R-matrix radius $\rho_m$ to the matching radius of the three-body asymptotic wavefunction $\rho_a$ is divided into sectors. We choose the size $h_p$ of sector $p$ to be sufficiently small that the interaction within is considered to be constant $\lambda_{\gamma}(p)$ in each channel $\gamma$. We diagonalize $\tilde{V}_{\gamma\gamma''}$ by solving the equation:
\begin{equation}
({\bf \tilde{T}}^p)^T\:{\bf \tilde{V}}(\rho_p)\:{\bf T}^p=\lambda(p)^2\, ,
\label{diag-eq}
\end{equation}
where $\rho_p$ is taken at the center of sector $p$. The wavefunction and its derivative at the left and right boundary of sector $p$ can be related through a local diagonal representation of the propagating functions ${\bf G}^p$:
\begin{equation}
\left[ \begin{array}{c} {\bm \chi}_R^p \\{\bm \chi}_L^p \end{array} \right] = \begin{bmatrix} {\bf G}_1^p & {\bf G}_2^p \\ {\bf G}_3^p & {\bf G}_4^p \end{bmatrix}  \left[ \begin{array}{c} -{{\bm \chi}^p_R}'\\ {{\bm \chi}^p_L}'\end{array} \right]\, .
\label{propa1-eq}
\end{equation}
The subscripts $R$, and $L$ imply evaluations at the right and left side of the sector boundary, respectively. The propagating functions are expressed as ${{\bf G}_i}^p={\bf T}^p{{\bf g}_i}^p{\bf \tilde{T}}^p$, where ${\bf g}^p$ is a simple function of $\lambda_{\gamma}(p)$:
\begin{eqnarray}
&&(g_1^p)_{\gamma\gamma'}=(g_4^p)_{\gamma\gamma'}=\delta_{\gamma\gamma'}
\begin{cases}
-\frac{1}{|\lambda_{\gamma}|}\:\mathrm{coth}|h_p\lambda_{\gamma}|,\quad \lambda_{\gamma}^2 > 0\\
\frac{1}{|\lambda_{\gamma}|}\:\mathrm{cot}|h_p\lambda_{\gamma}|,\quad \lambda_{\gamma}^2 \le 0\\
\end{cases}\nonumber\\
&&(g_2^p)_{\gamma\gamma'}=(g_3^p)_{\gamma\gamma'}=\delta_{\gamma\gamma'}
\begin{cases}
-\frac{1}{|\lambda_{\gamma}|}\:\mathrm{csch}|h_p\lambda_{\gamma}|,\quad \lambda_{\gamma}^2 > 0\\
\frac{1}{|\lambda_{\gamma}|}\:\mathrm{csc}|h_p\lambda_{\gamma}|,\quad \lambda_{\gamma}^2 \le 0\nonumber\\
\end{cases}\, .\\
\label{profunc-eq}
\end{eqnarray}
The R matrix is then propagated from sector $p{-}1$ to the next by:
\begin{equation}
{\bf R}^p=\frac{1}{\rho_R^{p}}\:\left({\bf G_2}^p\:[{\bf G_4}^p-\rho^{p-1}_R\:{\bf R}^{p-1}]^{-1}\:{\bf G_3}^p-{\bf G_1}^p\right)\, .
\label{rpropa-eq}
\end{equation}
Eq.~(\ref{rpropa-eq}) enables us to propagate the R matrix from a small radius $\rho_m$ to a distance $\rho_a$ large enough such that the asymptotic matching can be performed and hence the wavefunction at the R-matrix boundary can be calculated. Then the propagating functions ${\bf G}^p$ are used again to reconstruct the continuum wavefunctions to the desired radius 
\begin{eqnarray}
&&{\bm \chi}_L^p={\bf G_1}^{p-1}\:[{\bf G_3}^{p-1}]^{-1}\:{\bm \chi}_L^{p-1} \nonumber\\
&&+\left({\bf G_2}^{p-1}-{\bf G_1}^{p-1}\:[{\bf G_3}^{p-1}]^{-1}\:{\bf G_4}^{p-1}\right)\:{{\bm \chi}_L^{p-1}}'\, .
\label{wfspropa-eq}
\end{eqnarray}
\indent In our calculations, the logarithmic derivative $\beta$ is always kept at zero so that Eq.~(\ref{rpropa-eq}) and Eq.~(\ref{wfspropa-eq}) remains valid.
\subsection{Screening}
\begin{figure}
\center
\includegraphics[width=0.45\textwidth]{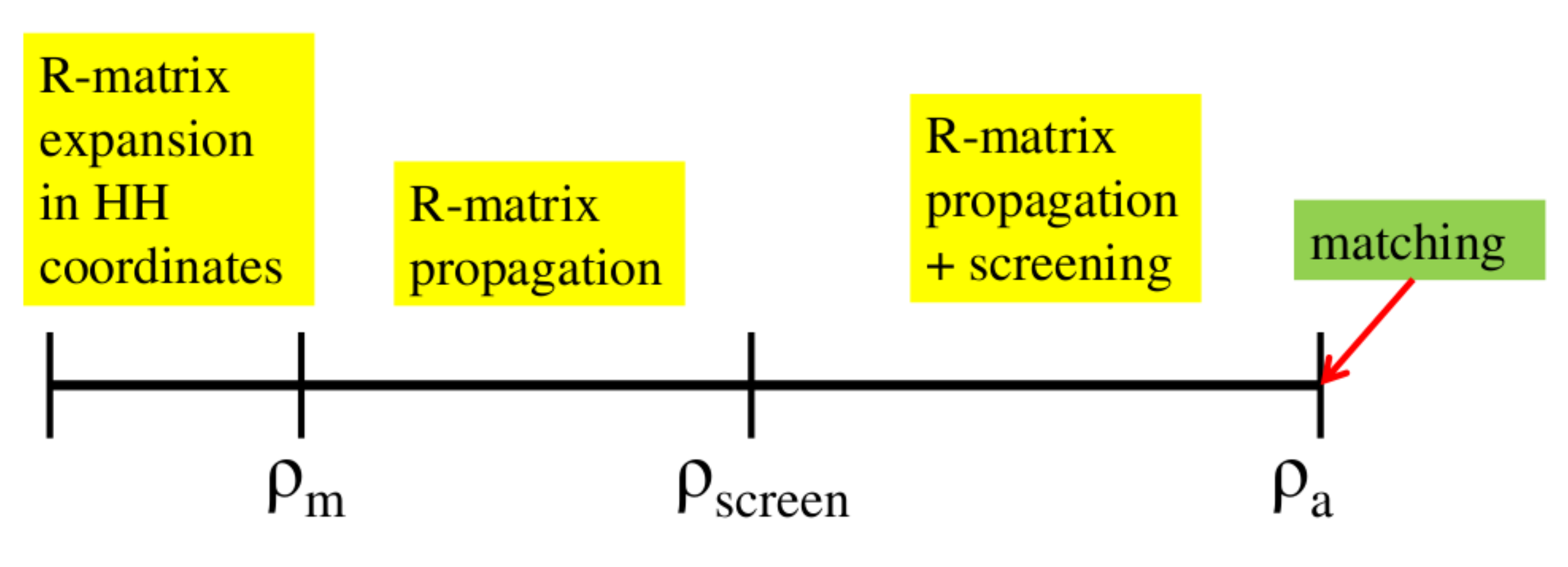}
\caption{(Color online) Our three-body method is divided into four steps in which we employ the R-matrix expansion, R-matrix propagation and screening technique in the Hyperspherical Harmonics coordinates.}
\label{region}
\end{figure}

The presence of very narrow two-body and three-body resonances in addition to the strong, long-range Coulomb interaction in our problem creates numerical instabilities in the propagated three-body scattering wavefunctions. We overcome these difficulties by introducing a Woods-Saxon screening factor $[1+\exp((\rho-\rho_{\mathrm{screen}})/a_{\mathrm{screen}})]^{-1}$ in the off-diagonal potentials of Eq.~(\ref{coupl-eq}). The screening radius $\rho_{\mathrm{screen}}$ is chosen sufficiently large to gain convergence, yet small enough to ensure numerical stability.\\
\indent Fig.~\ref{region} summarizes the different steps in our method. We divide the hyper-radial space into four regions. The Hyperspherical Harmonics R-matrix expansion (HHR) is first applied in a small box of radius $\rho_m$ ($\sim 50$ fm). We then use the R-matrix propagation method to much larger distances. All Coulomb couplings are included from $\rho_m$ to $\rho_{screen}$ ($\sim 800$ fm). After that, the off-diagonal couplings are screened up to $\rho_a$ ($\sim 3000$ fm) where it is safe to perform the asymptotic matching.
\subsection{Three-body reaction rate}
For a given three-body radiative capture reaction $a+b+c \rightarrow D + \gamma$, the reaction rate $R_{abc}$ at a given energy $E$ is calculated through the photo-dissociation cross section $\sigma_{\gamma}$
\begin{align}
R_{abc}(E)&= p!\: N_A^2 \; G_{abc,D} \; \frac{\hbar^3}{c^2}\:\frac{8\pi}{\left(\mu_{ab}\mu_{ab,c}\right)^{3/2}}
 \; \frac{E^2_{\gamma}}{E^2} \; \sigma_{\gamma}(E_{\gamma})\, .
\label{rate-eq}
\end{align}
The photon energy $E_{\gamma}$ and the three-body kinetic energy $E$ are related by $E_{\gamma}=E+|E_D|$. Here $E_D$ indicates the bound state energy of the final nucleus $D$ relative to the three incident-particle $a$, $b$, $c$ threshold. $\mu_{ab}$ and $\mu_{ab,c}$ are the reduced
masses of  the $a+b$ and $(ab)+c$ systems, respectively. $p$ is the number of identical particles and $N_A$ is Avogadro's number.  The statistical factor $G_{abc,D}$ depends on the spins of the nuclei $a$, $b$, $c$ and $D$ through the relationship $G_{abc,D}=\frac{2 (2J_D+1)}{(2J_a+1) (2J_b+1) (2J_c+1)}$.\\
\indent The energy averaged reaction rate $\langle R_{abc}\rangle(T)$ (a function of temperature) is an important quantity in astrophysics. It is obtained by integrating $R_{abc}(E)$ over the three-body Maxwell-Boltzmann distribution which differs from the two-body case:
\begin{equation}
\langle R_{abc}\rangle=\frac{1}{2}\:\frac{1}{(k_{B}T)^3}\:\int_{0}^{\infty}R_{abc}\:E^2\:e^{-\frac{E}{k_BT}}\:\mathrm{d}E \;,
\label{rateavg-eq}
\end{equation}
where $k_B$ is the Boltzmann constant. Throughout this paper, we will refer to the energy averaged reaction rate $\langle R_{abc}\rangle(T)$ as the triple-alpha rate.\\
\indent The photo-dissociation cross section $\sigma_{\gamma}$ in Eq.~(\ref{rate-eq}) depends on the electromagnetic transition strength $\mathrm{d}B(E\lambda)/\mathrm{d}E$ through which the reaction occurs:
\begin{equation}
\sigma_{\gamma}=\frac{(2\pi)^3(\lambda {+}1)}{\lambda [(2\lambda{+}1)!!]^2}\:\left(\frac{E_{\gamma}}{\hbar c}\right)^{2\lambda -1}\:\frac{\mathrm{d}B(E\lambda)}{\mathrm{d}E}\, .
\end{equation}
At low energies, the triple-alpha reaction proceeds through a quadrupole transition from the $0^+$ continuum to the $2^+_1$ bound state in $^{12}$C. The formula in the hyperspherical basis for the transition strength function $\mathrm{d}B(E2)/\mathrm{d}E$ is given in the Appendix. A detailed derivation of the three-body quadrupole transition and reaction rate is presented in \cite{thesis}.

\section{Results}
A new code named HHR3a is generated to solve Eq.~(\ref{hh-eq}) for bound states and continuum states of the triple-alpha system. It is developed from the program FaCE \cite{face} and STURMXX \cite{sturmxx} which are originally designed by I.J. Thompson {\it et al.} for a $core+n+n$ system. Our current problem, involving three identical alpha particles, has different symmetries and is more difficult because of the long-range Coulomb interaction. An alpha particle is considered as a boson of spin zero, thus the wavefunction must be unchanged for any permutation of these two particles. Only partial waves with $l_x=even$ will contribute to the total wavefunction. Here, $l_x$ is the relative angular momentum of the two identical alpha particles being interchanged. This constraint is applied for the wavefunction in each of the three Jacobi sets. Because the wavefunction is symmetric for any pair of interchanged particles, we are able to reduce three Faddeev components to a set of coupled channel equations in the hyper-radius formed from one pair of Jacobi coordinates. We thus highly increase the computational efficiency and fully take into account the symmetrization of the system. The new code HHR3a is corrected for the symmetry properties introduced by the three identical bosons. In addition, the correct Coulomb asymptotic wavefunctions are used instead of plane waves in STURMXX \cite{sturmxx} since our problem involves charged particles. We implement the screening technique in this new version of the code and considerable modifications are also introduced in the R-matrix propagation method. 
\subsection{Interactions \label{interaction}}
In our study, the triple-alpha problem is reduced to an effective three-body problem in which the two-body alpha-alpha interactions are phenomenological and a three-body force is needed to account for the fact that the alpha particle is not a fundamental particle. We use the Ali-Bodmer potential for the alpha-alpha interaction \cite{ali66} as modified by Fedorov {\it et al.} \cite{fedorov96}:
\begin{equation}
V_{\alpha\alpha}=(125\hat{P}_{l=0}+20\hat{P}_{l=2})e^{-r^2/1.53^2}-30.18e^{-r^2/2.85^2}.
\end{equation} 
The $l$-dependent potential has a very strong repulsive core in the s-wave to simulate the Pauli exclusion principle in the $\alpha-\alpha$ system. This interaction reproduces successfully the low energy phase shifts as well as the $0_1^+$ resonant state of $^{8}$Be at $0.093$ MeV. \\
\indent A three-body force is defined by $V_{3b}(\rho)=V_0e^{-\rho^2/\rho_0^2}$ and added to the Hamiltonian to reproduce the experimental bound state energy of the $2_1^+$ state as well as the Hoyle resonant state $0_2^+$ in $^{12}$C. We take $\rho_0= 6$ fm as it corresponds to the position where the three alpha particles touch each other. We adjust $V_0$ to the value of $-15.94$ MeV ($-19.46$ MeV) to generate the $2_1^+$ bound state (the $0_2^+$ Hoyle resonance) at the correct binding energy. $V_0=-19.46$ MeV is then used to calculate all the $0^+$ continuum states. \\
\indent In addition to the nuclear interaction we include the Coulomb potential:
\begin{equation}
V^{Coul}_{\alpha\alpha}(r)=Z^2e^2\times\left\{ 
  \begin{array}{l l}
   \left(\frac{3}{2}-\frac{r^2}{2r^2_{Coul}}\right)\frac{1}{r_{Coul}} & \quad \text{$r\le r_{Coul}$}\\
   \frac{1}{r} & \quad \text{$r \ge r_{Coul}$\, .}\\
  \end{array} \right.\label{coul-eq}
\end{equation}
Here, $Z$ is the charge of an alpha particle, $r$ is the distance between two alpha particles, and $r_{Coul}$ is the Coulomb radius which is taken as twice the alpha particle radius $r_{Coul}=2.94$ fm.\\
\indent In Table \ref{t1}, we summarize the parameters for the three-body interactions used in our calculations to reproduce the experimental binding energy for the $2^+_1$ bound state and the resonant energy for the Hoyle state $0^+_2$ of $^{12}$C \cite{ajzenberg90}.
\begin{table}[b!]
\caption{The three-body interactions to reproduce the experimental energies \cite{ajzenberg90} of the $2^+_1$ bound state and $0^+_2$ resonant state in $^{12}$C}
\label{t1}
\begin{ruledtabular}
\begin{tabular}{cccc}
$J^{\Pi}$  &$V_0$(MeV) & $\rho_0$(fm) & $E $(MeV)\\ \hline
$2^+_1$ & -15.94 & 6 &-2.875 \\
$0^+_2$ & -19.46 & 6 & 0.380 \\
\end{tabular}
\end{ruledtabular}
\end{table}
\subsection{The $2^+_1$ bound state \label{2bs}}
\begin{figure}
\center
\includegraphics[width=0.45\textwidth]{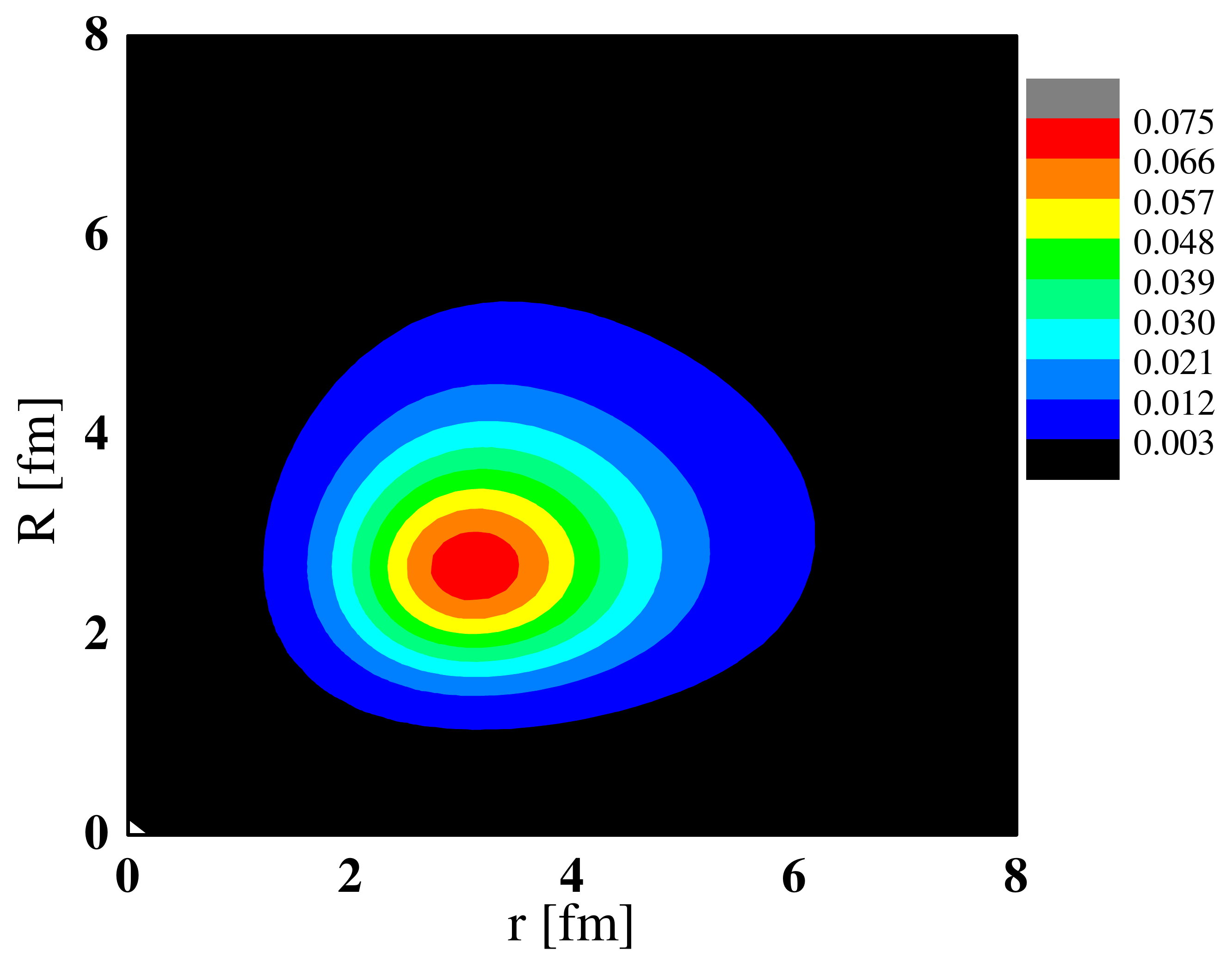}
\caption{(Color online) The density distribution for the $2_1^+$ bound state of $^{12}$C. The brighter color corresponds to the higher density distribution.}
\label{2bsdens}
\end{figure}

A three-body bound state is obtained by solving the set of coupled channels equations (\ref{hh-eq}) with a boundary condition that the wavefunction goes to zero at large $\rho$. These coupled equations are solved by expanding the hyper-radial wavefunction in terms of the Laguerre basis \cite{erens71} which forms a complete and orthogonal set. We use a large basis to obtain an accurate description at large distances ($n_{lag}=180$). The matrix diagonalization is then applied to obtain eigenenergies and eigenvectors. For the hyperangular part we use $n_{jac}=70$ Jacobi polynomials in the expansion.  We use the modified Ali-Bodmer interaction \cite{ali66, fedorov96} in combination with a three-body force as discussed in Sec.~\ref{interaction} to obtain the $2_1^+$ bound state for $^{12}$C in agreement with experiment. \\ 
\indent
\begin{figure}
\center
\includegraphics[width=0.45\textwidth]{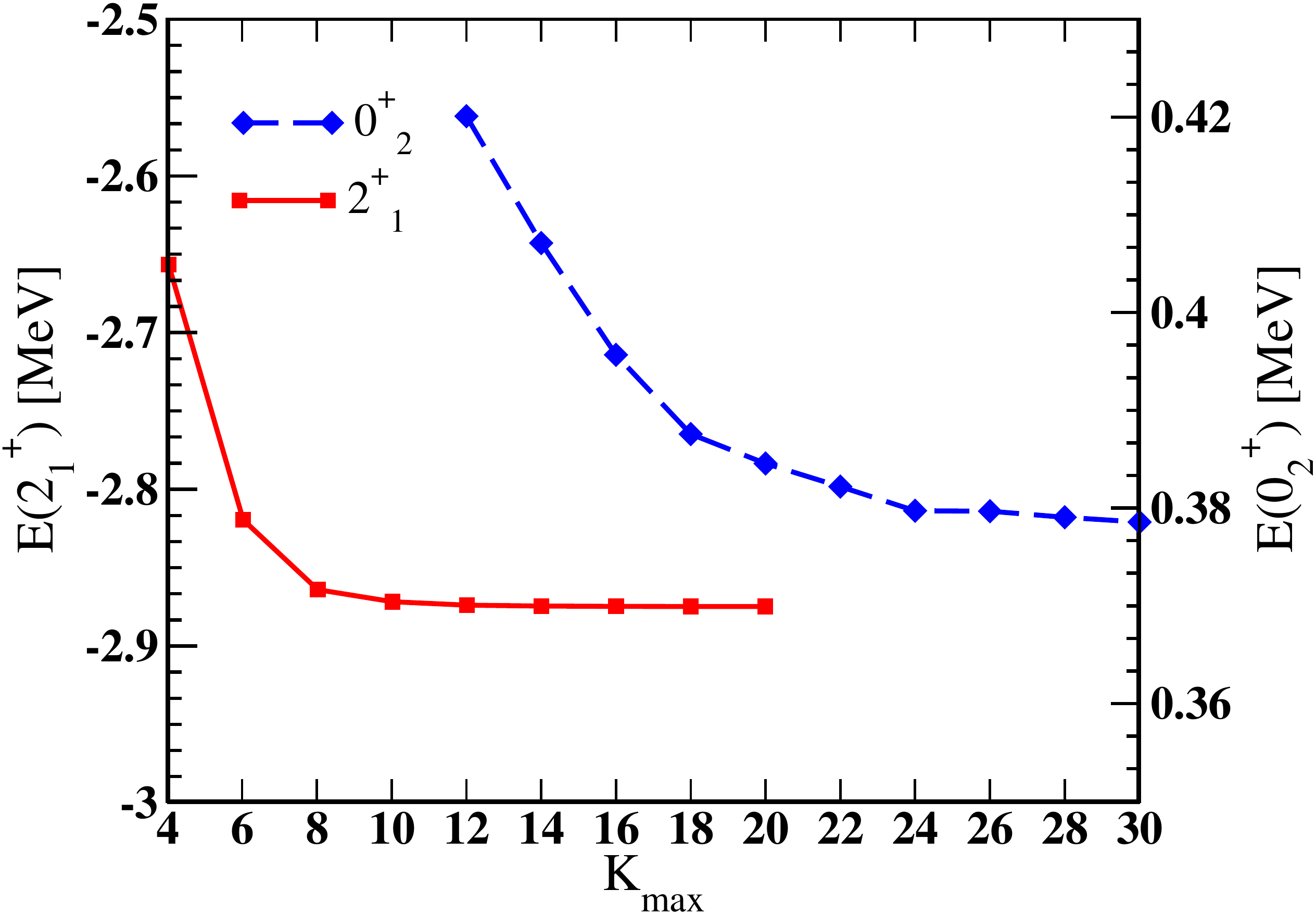}
\caption{Dependence of the three-body binding energy of the $^{12}$C($2_1^+$) bound state and the $^{12}$C($0_2^+$) Hoyle resonant state on the maximum hyper-momentum $K_{\max}$ which determines the size of the model space. Any $K \leq K_{\max}$ will be included in the wavefunction expansion Eq.~(\ref{hhexpan-eq}).}
\label{f1}
\end{figure}
\begin{figure}
\center
\includegraphics[width=0.45\textwidth]{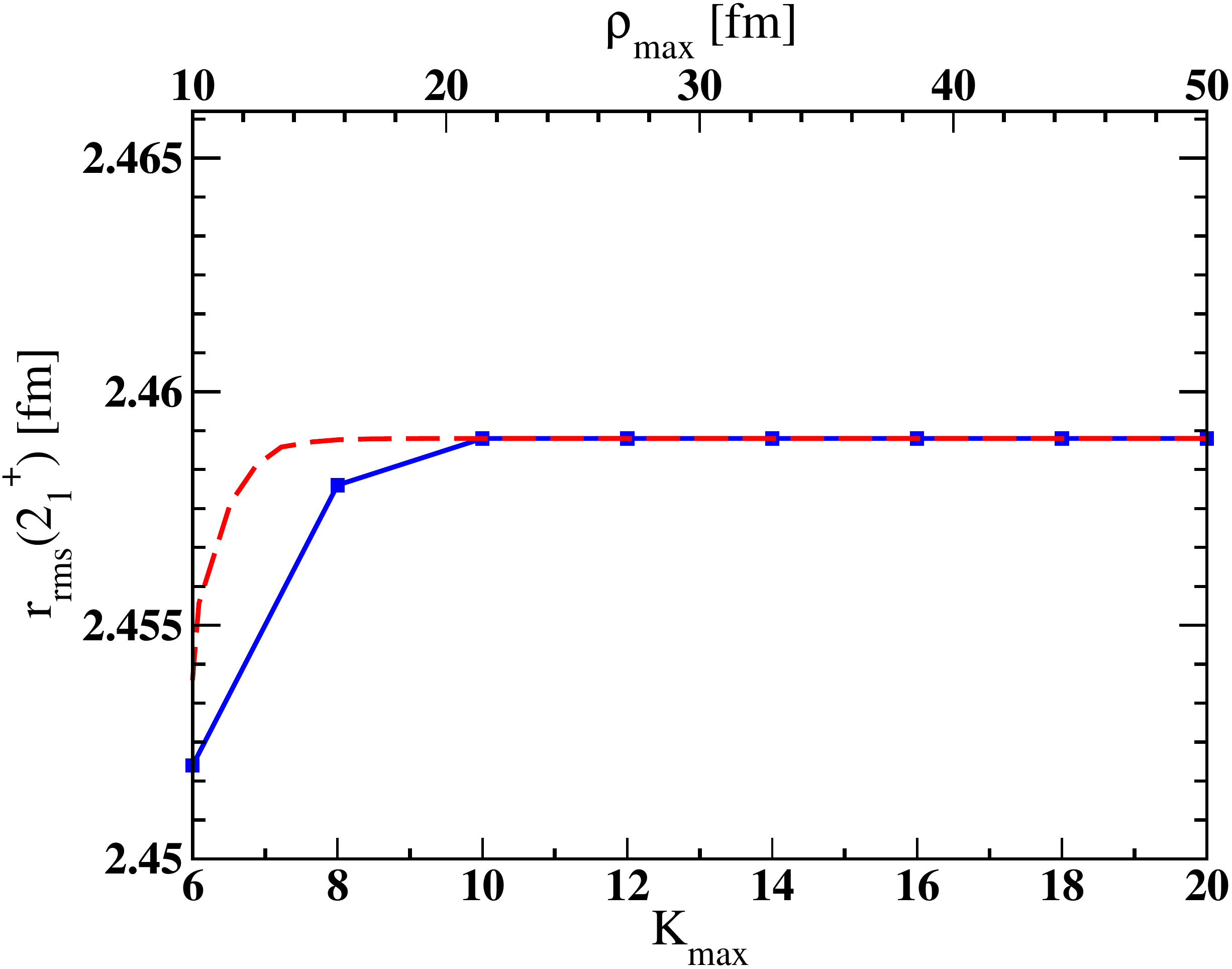}
\caption{Dependence of the $r_{\mathrm{rms}}$ radius of the $^{12}$C($2_1^+$) bound state on the maximum hyper-momentum $K_{\max}$ (square-solid) and the maximum radius of the calculated $^{12}$C($2_1^+$) wavefunction (dashed).} 
\label{f2}
\end{figure}
Fig.~\ref{f1} and Fig.~\ref{f2} show the convergence of the three-body binding energy and the rms radius of the $2_1^+$ bound state of $^{12}$C with the size of the model space represented here by the maximum hyper-momentum $K_{\max}$. A given $K_{\max}$ determines how many channels are included in the wavefunction expansion Eq.~(\ref{hhexpan-eq}), e.g., $K_{\max}=20$ produces 36 channels in the expansion. From Fig.~\ref{f1} (square-solid curve), we see that the bound state energy starts converging at $K_{\max}=12$. Therefore, energy convergence is guaranteed by choosing $K_{\max}=20$. The bound state energy of the $2_1^+$ state of $^{12}$C converges to the experimental value $E=-2.875$ MeV \cite{ajzenberg90} with respect to the three-alpha threshold. Our final value for the rms radius is $2.459$ fm (see Fig.~\ref{f2}, square-solid curve) which is close to the result obtained in a microscopic calculation done by Chernykh {\it et al.} ($2.50$ fm) \cite{chernykh07}. This agreement validates the three-body approximation for the $2_1^+$ bound state. In support of this is also the fact that in \cite{chernykh07}, the triple-alpha configuration is dominant for this  $^{12}$C state.\\
\indent {In Fig.~\ref{f2} (dashed curve), we plot the $r_{\mathrm{rms}}$ radius as a function of the maximum radius $\rho_{\mathrm{max}}$ of the calculated $2_1^+$ bound state wavefunction. As can be seen, the $r_{\mathrm{rms}}$ no longer depends on the maximum radius of the bound state wavefunction for $\rho_{\mathrm{max}} > 20$ fm. However this is not generally the case when considering capture rates. All the calculations throughout this work employ a bound state wavefunction up to $200$ fm, which ensure the correct description of the long-range part of the wavefunction important for the rate at low temperatures.}\\
\indent We next construct the density distribution function from the $2^+$ bound state wavefunction to study its structure:
\begin{equation}
P(r,R)=\int\left|\Psi({\bf r},{\bf R})\right|^2\mathrm{d}\hat{{\bf r}}\mathrm{d}\hat{{\bf R}} \, .
\label{dens-eq}
\end{equation}
Here $r$ is the radius between two alpha particles and $R$ is the distance from their center of mass to the third alpha. In Fig.~\ref{2bsdens} we present the density distribution $P(r,R)$ for the $2_1^+$ bound state of $^{12}$C. The dominant configuration for this state is an equilateral triangle in which each pair of particles is $\sim 3$ fm apart. This configuration is represented in Fig.~\ref{config}c.
\subsection{The $0^+_2$ Hoyle state}
\begin{figure}
\center
\includegraphics[width=0.45\textwidth]{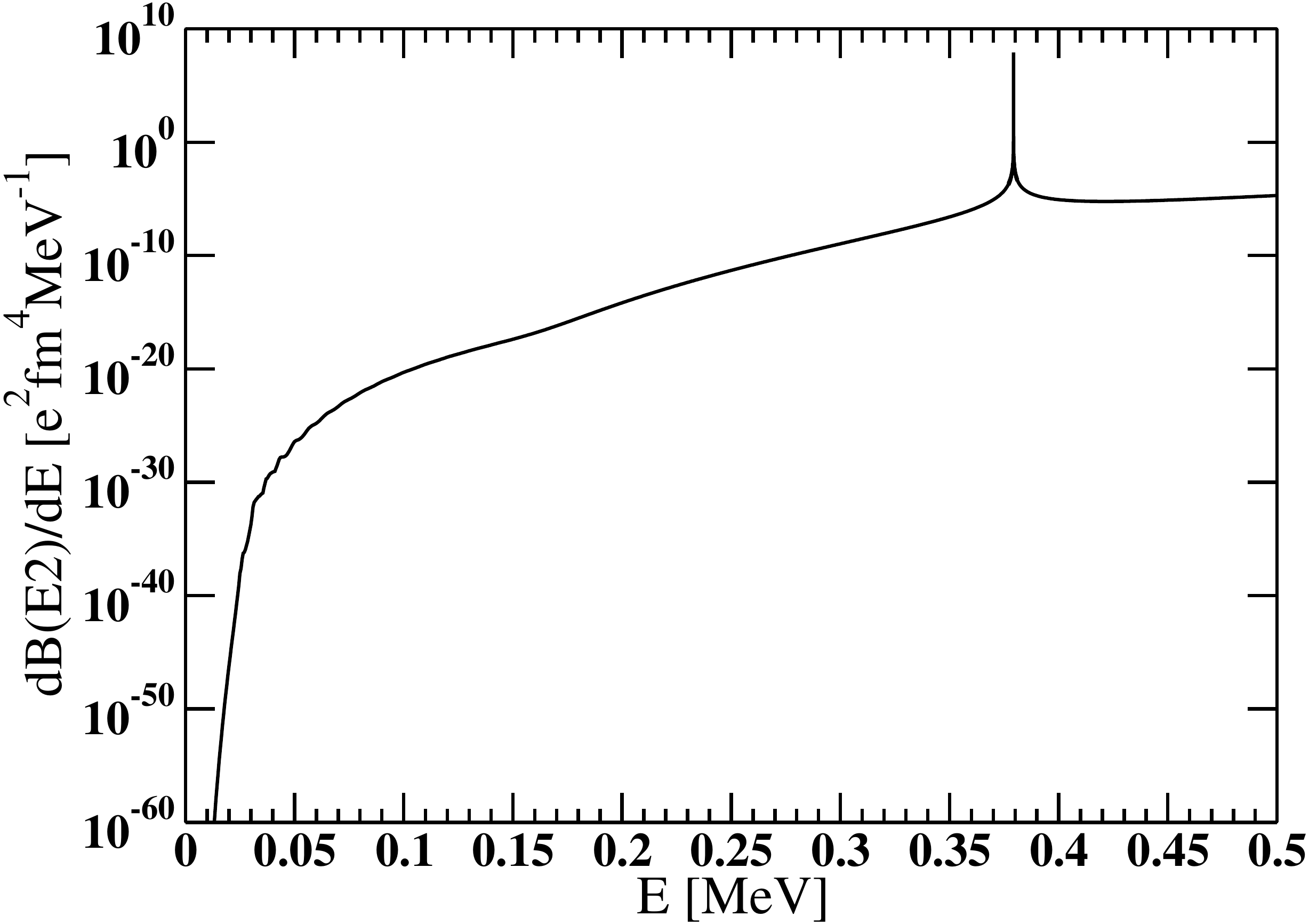}
\caption{The quadrupole transition strength $\mathrm{d}B(E2)/\mathrm{d}E$ as a function of the three-alpha kinetic energy $E$.}
\label{f4}
\end{figure}

\begin{figure}
\center
\includegraphics[width=0.45\textwidth]{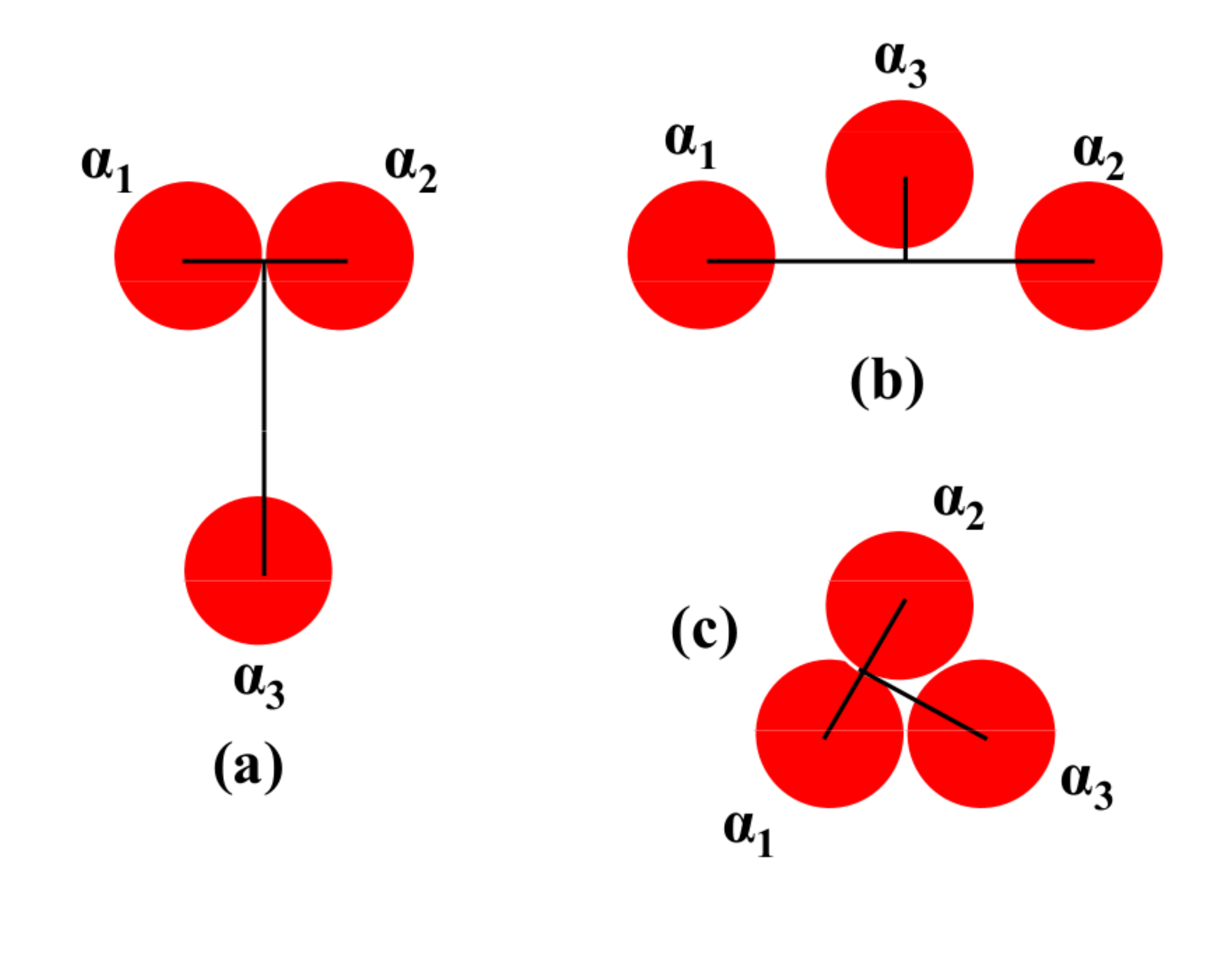}
\caption{(Color online) Different configurations of the triple-alpha system: the prolate triangle (a), the oblate triangle (b), and the equilateral triangle (c).}
\label{config}
\end{figure}

The three alpha particle system is driven by a strong Coulomb interaction of which the off-diagonal couplings are long-range in the HH representation. Therefore, solving the coupled channels equations (\ref{hh-eq}) for positive energies $E$ is not a trivial task. The HHR method which combines the R-matrix expansion and the R-matrix propagation in the HH basis is employed to overcome this difficulty. We again use the modified alpha-alpha Ali-Bodmer potential \cite{ali66, fedorov96} and adjust the three-body force to reproduce the experimental Hoyle resonant state (details of interactions are shown in Sec.~\ref{interaction}). The same three-body force is then used to calculate all the $0^+$ continuum states of $^{12}$C. The quadrupole transition strength Eq.~(\ref{be2e-eq}) is constructed using the $2^+_1$ bound state wavefunction and the $0^+$ resonant and non-resonant continuum states. Therefore, in our calculations we treat the resonant and non-resonant process on the same footing. Fig.~\ref{f4} plots the quadrupole transition strength as a function of the kinetic energy for the three interacting alpha particles. The curve peaks at the measured Hoyle resonant energy ($E=0.38$ MeV). As we expect, the strength function decreases as it approaches the lower energy regime. Around $0.05$ MeV there is a sharp reduction of the transition strength below which the formation of $^{12}$C becomes unlikely.\\
\indent We repeated the calculation in Fig.~\ref{f4} for different values of $K_{\max}$. For each $K_{\max}$, the resonant energy is obtained from the maximum of the transition strength function and plotted in Fig.~\ref{f1} (diamond-dashed curve). As expected, we find that the convergence for the Hoyle state is slower than for the bound state. The result begins to converge at $K_{\max}=26$ and this is the point we choose to fit the three-body force to reproduce the experimental energy of $0.38$ MeV \cite{ajzenberg90} with respect to the three-alpha threshold. The resonant energy in Fig.~\ref{f1} has a typical exponential convergence pattern with the maximum hyper-momentum $K_{\max}$ in the HH coordinates. In order to estimate the uncertainly in extracting this resonant energy, the data in Fig.~\ref{f1} is fitted to an exponential function $y=A+Be^{-Cx}$. When $K_{\max}$ goes to infinity this function approaches its converged value which is then compared to the value of energy at $K_{\max}=26$ to obtain an uncertainty of $4\%$. \\ 
\begin{figure}
\center
\includegraphics[width=0.45\textwidth]{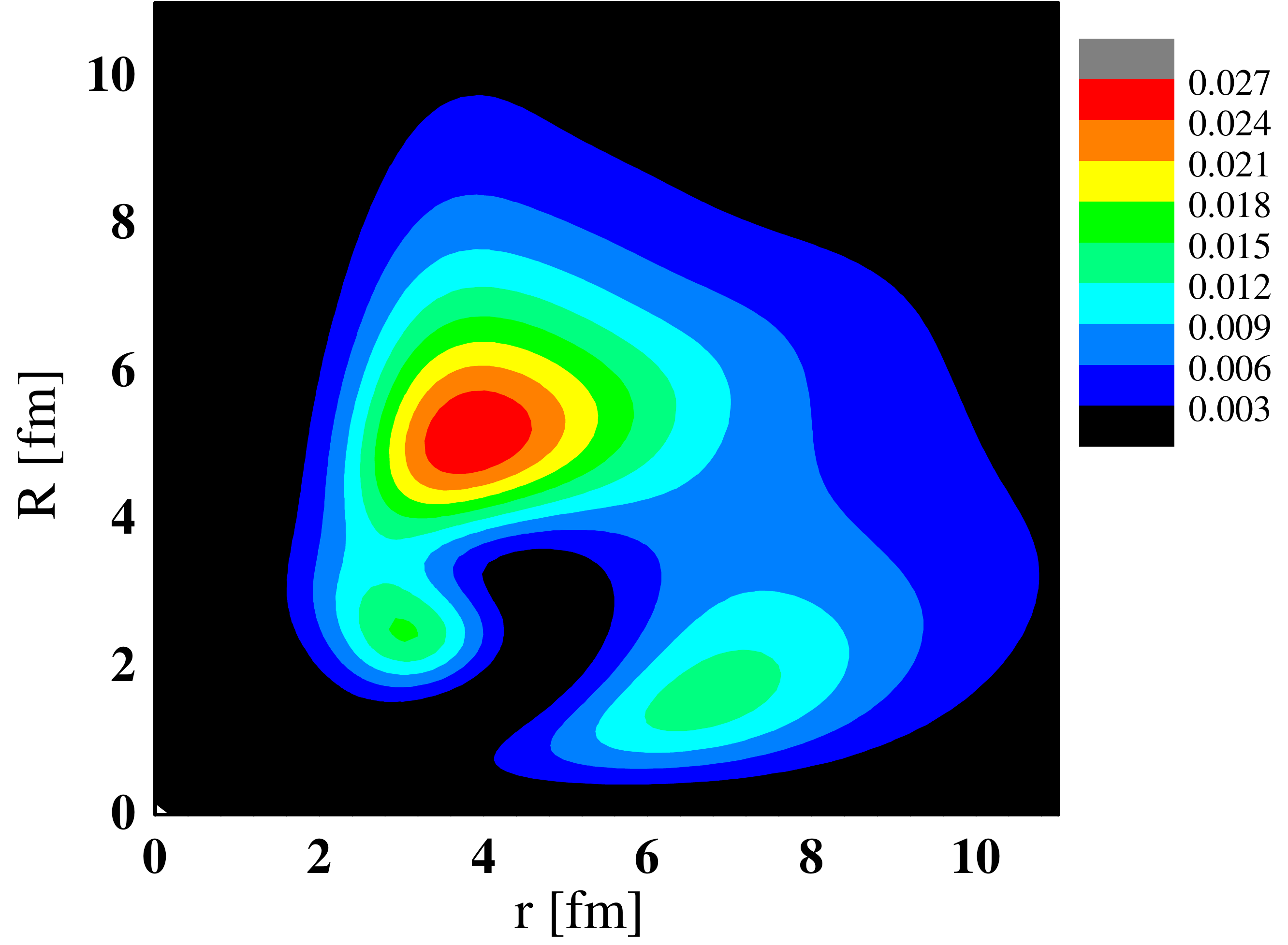}
\caption{(Color online) The density distribution for the $0_2^+$ Hoyle resonant state of $^{12}$C. The brighter color corresponds to the higher density distribution.}
\label{f5}
\end{figure}
\indent By using the wavefunction at the resonant energy we are able to construct a density distribution function as in Eq.~(\ref{dens-eq}). This quantity illustrates the spatial structure of the Hoyle state, a topic of trememdous interest for many years \cite{anagnostatos95,neff04,chernykh07,epelbaum11,vasilevsky12}. Fig.~\ref{f5} depicts the density distribution in a color contour plot where the brighter color corresponds to a denser distribution. The dominant configuration for the Hoyle state is the prolate triangle in which two alpha particles are near each other ($\sim 3.7$ fm) and further away from the third alpha ($\sim 5.2$ fm), in agreement with the findings in \cite{fedorov96, vasilevsky12} (presented in Fig.~\ref{config}a). We also observe two smaller maxima for the density distribution function. One of them supports an oblate triangle configuration of three alpha particles (Fig.~\ref{config}b), where two alpha particles are $\sim 7$ fm apart and their center of mass is $\sim 1.5$ fm from the third alpha. The other small maximum indicates an almost equilateral triangle with a distance of $\sim 3$ fm between any pair of alpha particles (Fig.~\ref{config}c). The weights of these two configurations are about half of that of the prolate triangle configuration. In conclusion, we find three different configurations in the Hoyle state: the prolate triangle is dominant, but the oblate triangle (almost chain-like) and the equilateral triangle also contribute significantly. Fig.~\ref{config} portrays accurately the configuration of three alpha clusters only in relative magnitudes of $r$ and $R$, and the average angle between these two vectors requires further investigation.
\subsection{Convergence and uncertainty \label{conv}}
\begin{figure}
\center
\includegraphics[width=0.45\textwidth]{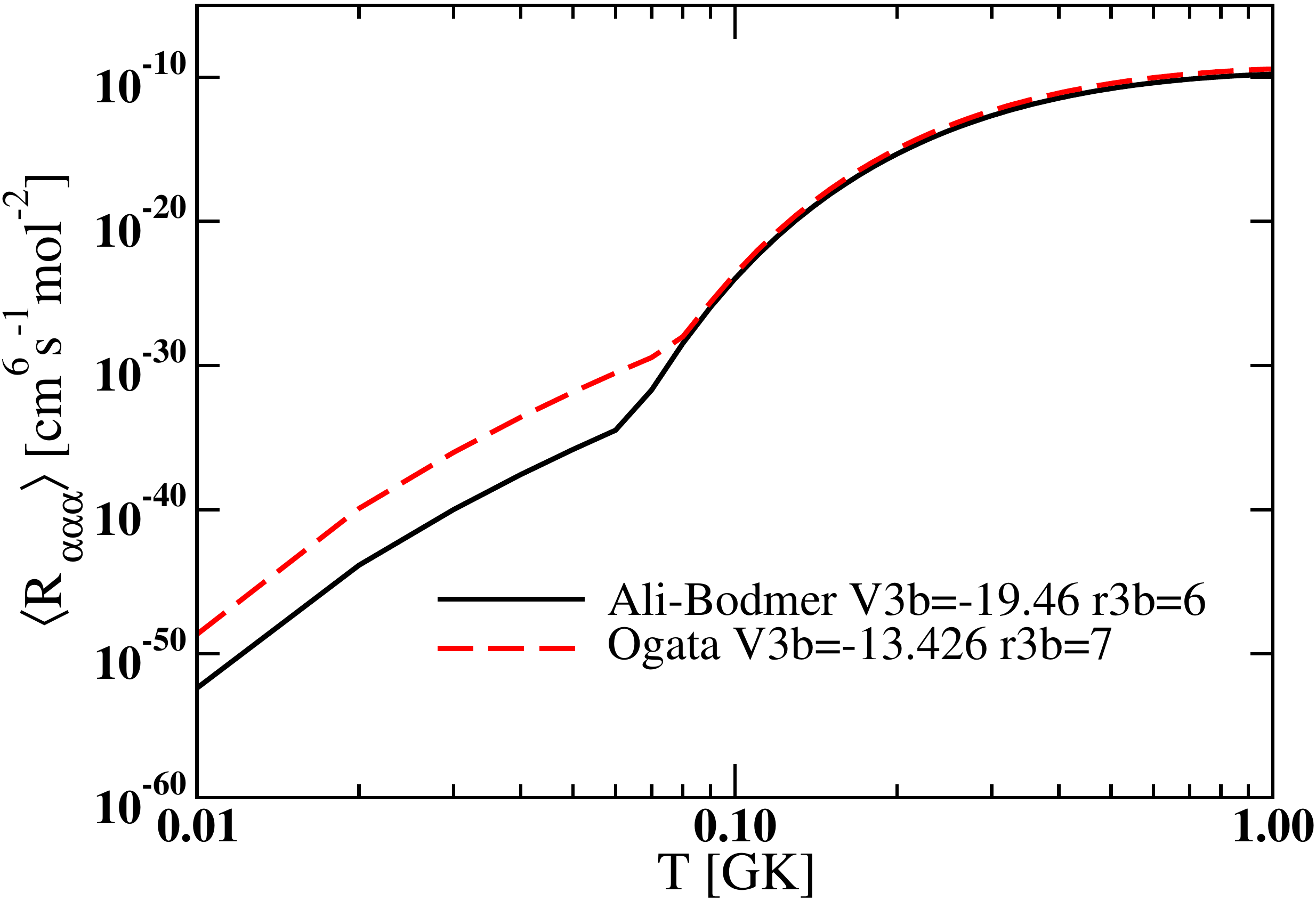}
\caption{(Color online) The sensitivity of the triple-alpha rate to the interactions: comparison between $\alpha-\alpha$ Ali-Bodmer+3-body force (solid) and $\alpha-\alpha$ interaction as in \cite{ogata09}+3-body force (dashed).}
\label{f13}
\end{figure}

It is important for us to carefully study the convergence properties of our method, given the well known difficulty of the problem. With the help of the R-matrix propagation method we are able to perform calculations out to a very large radius and obtain stable results with Coulomb interactions only. When nuclear interactions are introduced, the very narrow resonances occur in both the two-body and three-body systems, and lead to a degradation of our propagation technique and increasing numerical instabilities, especially in the turning point region. As mentioned in the theory section, we tackled this problem by screening the off-diagonal couplings by a Woods-Saxon multiplying factor $[1+\exp((\rho-\rho_{\mathrm{screen}})/a_{\mathrm{screen}})]^{-1}$. However, it is necessary for us to ensure that important physics is not left out. \\
\begin{figure}
\center
\includegraphics[width=0.45\textwidth]{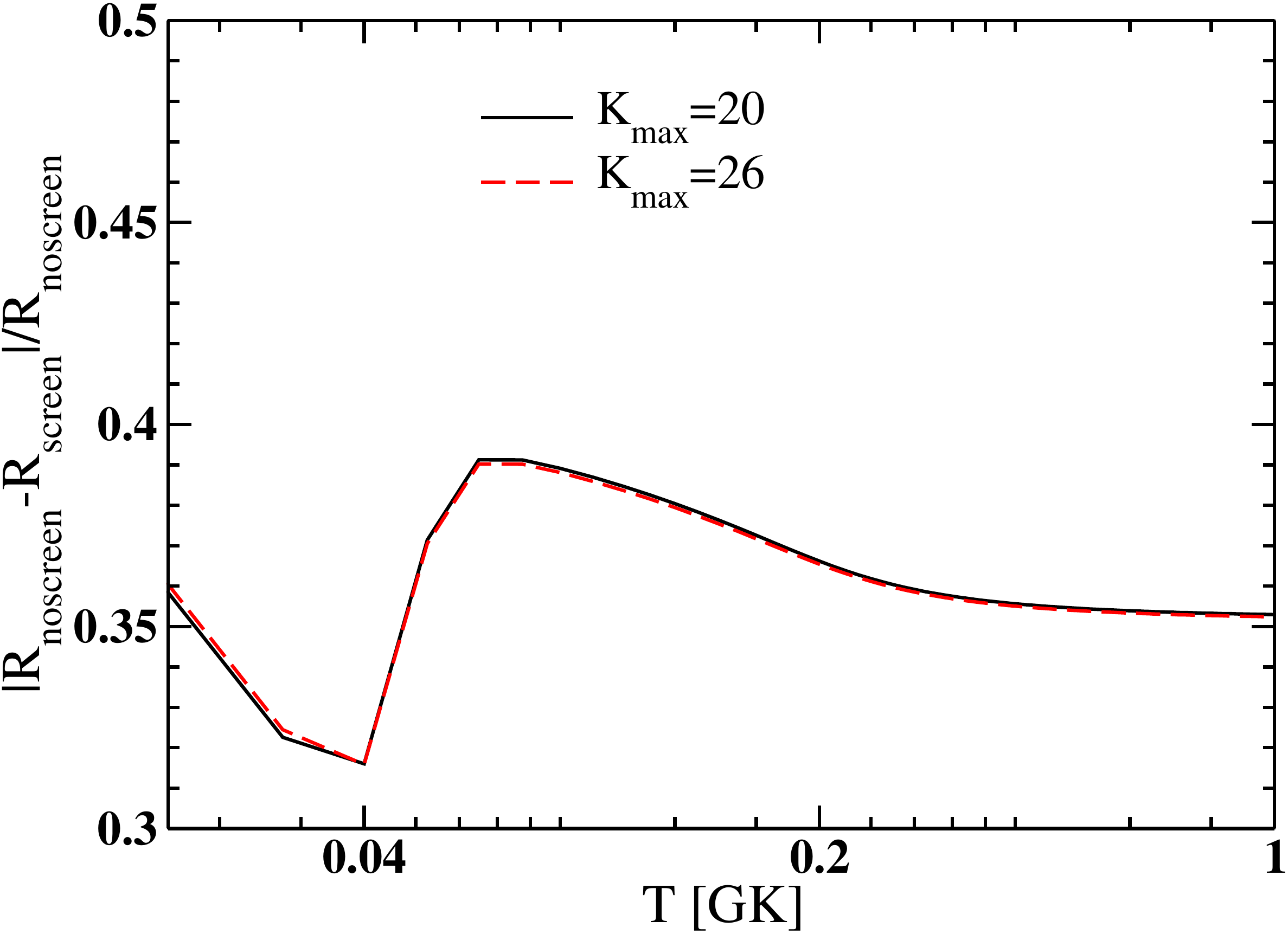}
\caption{(Color online) Relative differences between screening and no screening calculations for the fictitious three-alpha scattering system in which only Coulomb interactions are introduced. The screening calculation is peformed for $K_{\max}=26$ and $\rho_{\mathrm{screen}}=800$ fm. Calculations without screening are performed for $K_{\max}=20$ (solid) and $K_{\max}=26$ (dashed).}
\label{f6}
\end{figure}
\indent In Fig.~\ref{f6}, we compare between screening and no screening calculations for a scattering system of three alpha particles when only Coulomb interactions are included. The $2^+_1$ bound state is taken from Sec.~\ref{2bs}. For this case, there are no resonances in either the two-body or three-body system, thus the propagation technique is stable and we are able to determine the impact of introducing screening potentials. We perform a calculation using the screening technique with $\rho_{\mathrm{screen}}=800$ fm. This calculation fully converges at $K_{\max}=26$. This result is then compared to two calculations with no screening at all. One is done with $K_{\max}=20$ (solid) and the other with $K_{\max}=26$ (dashed) to ensure generality and the convergence of the method. Uncertainty of the reaction rate due to the screening technique ranges from $35\%$ to $39\%$ for temperatures within $0.02-1$ GK. These are very small numbers given the many orders of magnitude involved in the problem.\\
\begin{figure}
\center
\includegraphics[width=0.45\textwidth]{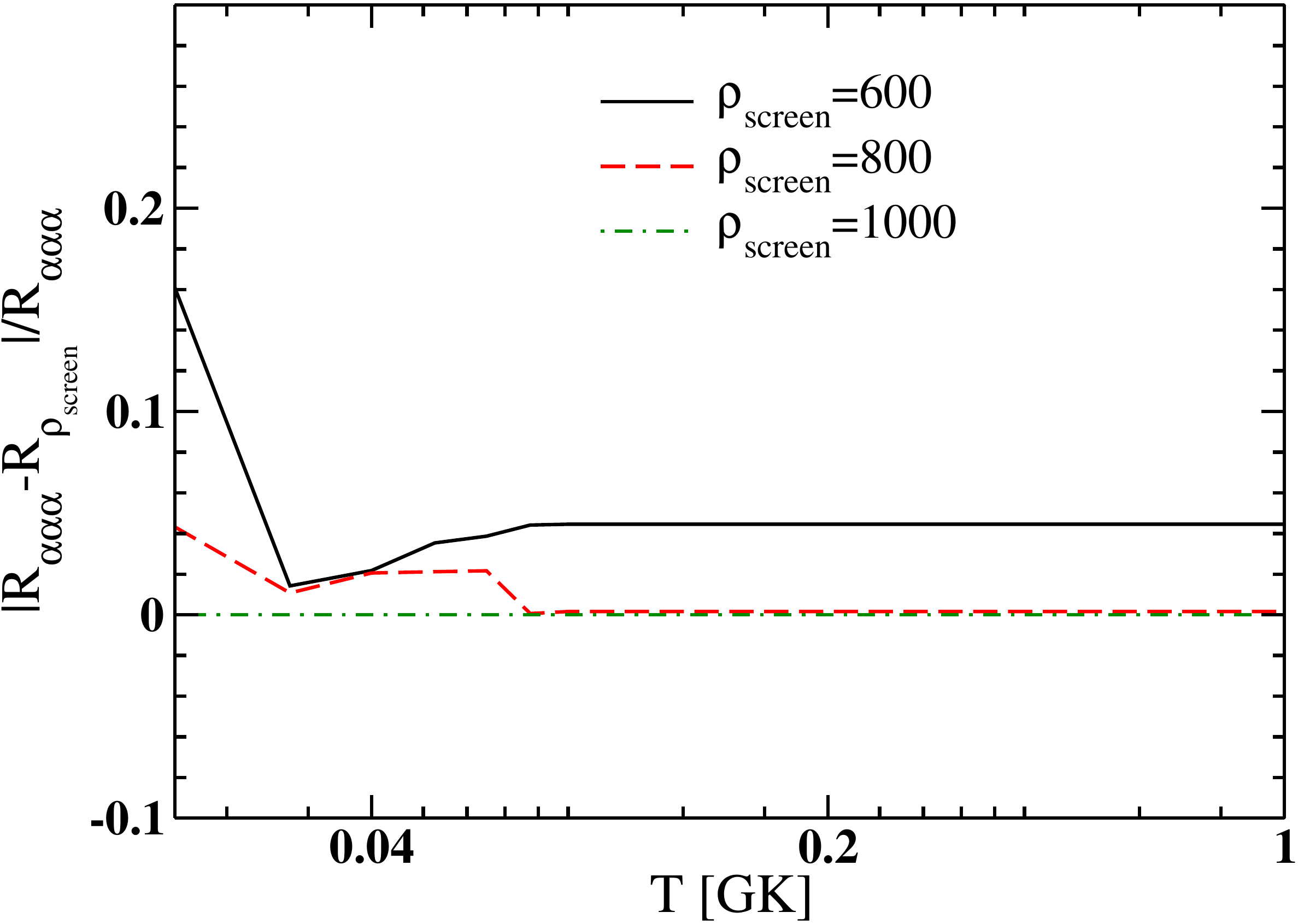}
\caption{(Color online) Relative differences between the triple-alpha rate calculated at several screening radii and its converged value. The reaction rates calculated using $\rho_{\mathrm{screen}}=600$ fm (solid), $\rho_{\mathrm{screen}}=800$ fm (dashed), $\rho_{\mathrm{screen}}=1000$ fm (dot-dashed) are compared with the converged rate in Table~\ref{t2}.}
\label{f7}
\end{figure}
\indent For our realistic problem of the triple-alpha where both nuclear and Coulomb interactions are included, we need to use the screening technique to obtain stable results. The convergence study of our triple-alpha rate with varying $\rho_{\mathrm{screen}}$ and $K_{\max}$ is performed carefully. We use an R-matrix box size of $50$ fm, with $P=50$ poles included in the R-matrix expansion. For the hyperangular part we use $n_{jac}=100$ Jacobi polynomials, which provide converged values for the hyperangular integrals. Several values of the screening diffuseness are used and no dependence has been found, we thus fix the screening diffuseness to $10$ fm throughout this paper. The R matrix is then propagated out to $3000$ fm, sufficiently large to perform the Coulomb asymptotic matching. Fig.~\ref{f7} presents the results using three different screening radii: 600, 800 and 1000 fm. These results are plotted in comparision with the converged rate in Table~\ref{t2}. The triple-alpha reaction rate converges for $\rho_{\mathrm{screen}}\ge 800$ fm with an uncertainty of $5\%$.\\
\begin{figure}
\center
\includegraphics[width=0.45\textwidth]{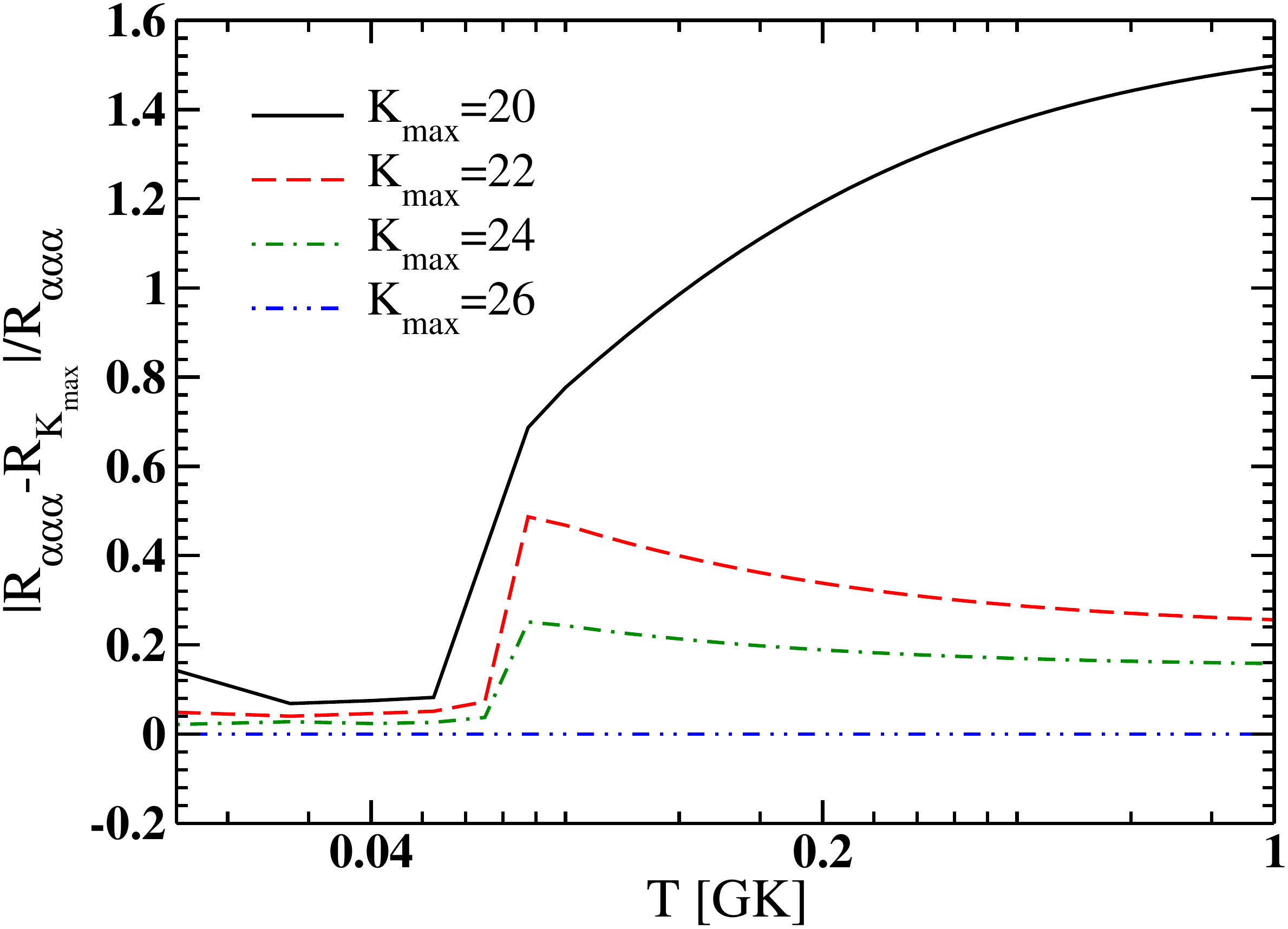}
\caption{(Color online) Relative differences between the triple-alpha rate calculated at several maximum hyper-momentum $K_{\max}$ and its converged value. The reaction rates calculated using $K_{\max}=20$ (solid), $K_{\max}=22$ (dashed), $K_{\max}=24$ (dot-dashed), $K_{\max}=26$ (dot-dot-dashed) are compared with the converged rate in Table~\ref{t2}. The calculation for $K_{\max}=20$ is rescaled by factor of 100 for $T>0.06$ GK.}
\label{f8}
\end{figure}
\indent Fig.~\ref{f8} shows the convergence of the triple-alpha reaction rate with hyper-momentum $K_{\max}$. We perform four calculations of the rate as a function of temperature by varying $K_{\max}$ from $20$ to $26$. The results are plotted in comparision with the converged rate in Table~\ref{t2}. $K_{\max}=20$ is not large enough to reproduce the experimental $0_2^+$ Hoyle resonant state, therefore the reaction rate for this case is larger than other curves for $T>0.1$ GK. The uncertainty of the triple-alpha rate is larger at high temperatures due to the sensitivity of the Hoyle resonant energy with $K_{\max}$. The uncertainty is less than $15\%$ for $K_{\max}=26$.\\
\begin{figure}
\center
\includegraphics[width=0.45\textwidth]{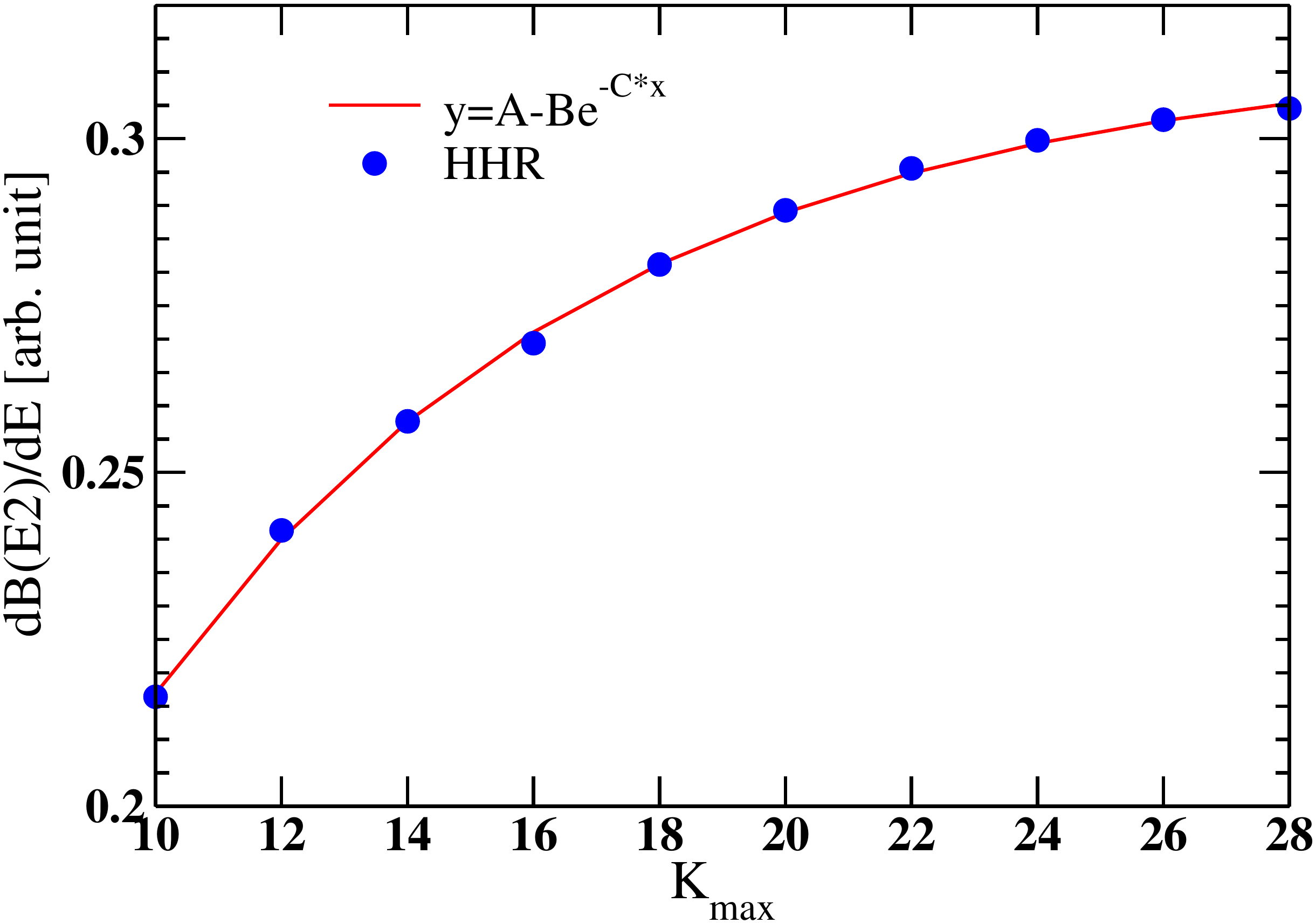}
\caption{(Color online) Convergence of the quadrupole transition strength $\mathrm{d}B(E2)/\mathrm{d}E$ with hyper-momentum $K_{\max}$ at $E=0.01$ MeV.}
\label{f9}
\end{figure}
\indent In Fig.~\ref{f9} we look at the quadrupole strength function, which is a principal ingredient for calculating the reaction rate, at a specific energy of $E=0.01$ MeV. If $\mathrm{d}B(E2)/\mathrm{d}E$ converges, then the cross section and the reaction rate will also converge. Since the $\mathrm{d}B(E2)/\mathrm{d}E$ has an exponential behavior, namely $y=A-Be^{-Cx}$ as a function of hyper-momentum $K_{\max}$, we are able to extrapolate to the converged value of $\mathrm{d}B(E2)/\mathrm{d}E$ when $K_{\max}\rightarrow\infty$. For this very low energy ($E=0.01$ MeV), the extrapolated value of the rate differs from that at $K_{\max}=26$ by only $4\%$. \\
\indent In order to study the sensitivity of the triple-alpha rate to the nuclear interaction and the three-body force, we consider employing the same Hamiltonian as \cite{ogata09} to calculate the $0^+$ continuum. In \cite{ogata09} the alpha-alpha interaction is given as: 
\begin{equation}
V_{\alpha\alpha}(r)=100.0e^{-r^2/1.00^2}-30.35e^{-r^2/2.13^2}\, .
\end{equation}
Unlike in \cite{ogata09}, we find that a three-body interaction is needed to reproduce the relevant Hoyle state. As seen from Fig.~\ref{f13}, the new interaction produces a triple-alpha reaction rate at low temperatures 4 orders of magnitude higher than that obtained with the Ali-Bodmer interaction \cite{ali66, fedorov96}. While the Ali-Bodmer potential reproduces the $\alpha\textrm{-}\alpha$ phase shifts, the interaction in \cite{ogata09} does not. It therefore provides an upper limit for the error associated with $V_{\alpha\alpha}$ ambiguities.\\
\indent As an additional check of our method, we have reproduced the results in \cite{descouvemont10b}(Table 4, second column). The aim of \cite{descouvemont10b} was the study of $3\alpha$ resonant structures up to 6 MeV, comparing results using shallow and deep $\alpha\alpha$ potentials. This is very different from our goal, which is to calculate the scattering at very low energies, below 0.3 MeV, where no resonances exist. Nevertheless, it is critical to ensure that our method does reproduce previous results. Using the same Ali-Bodmer interaction and the same three-body force, we reproduce the bound states exactly, and obtain three resonances: one at 1 MeV, another at 1.4 MeV and the last one at 3.3 MeV. The first and last correspond to the resonances at 0.93 MeV and 3.1 MeV in \cite{descouvemont10b}, while our 1.4 MeV is a broad resonance which is likely to correspond to the overlap of the three states identified in \cite{descouvemont10b}. We should point out that  these resonances do not exhibit the typical Breit-Wigner shape, and therefore the estimate of their widths is not very reliable. This is in contrast to the Hoyle state in our present work, where an isolated very narrow resonance is obtained. \\
\indent Our final rate is normalized to the standard rate from NACRE for temperatures $T\ge0.5$ GK, where the Hoyle resonance completely dominates. At this high temperature, the triple-alpha reaction proceeds through the very narrow Hoyle resonance and its rate is proportional to a gamma decay width $\Gamma$. In NACRE, $\Gamma=3.7\times 10^{-9}$ MeV is taken from experimental data \cite{ajzenberg90}. We fit our calculated cross sections to a Breit-Wigner shape and obtain a value of $\sim 10^{-9}$ MeV for the partial decay width, which is the same order of magnitude as experiment. In practice, a factor of 2 is needed for normalizing to the NACRE rate. We consider this factor of 2 to be our conservative error counting for both the uncertainty of the decay width and the convergence of the problem.
\subsection{Rate \label{rate}}
\begin{figure}
\center
\includegraphics[width=0.45\textwidth]{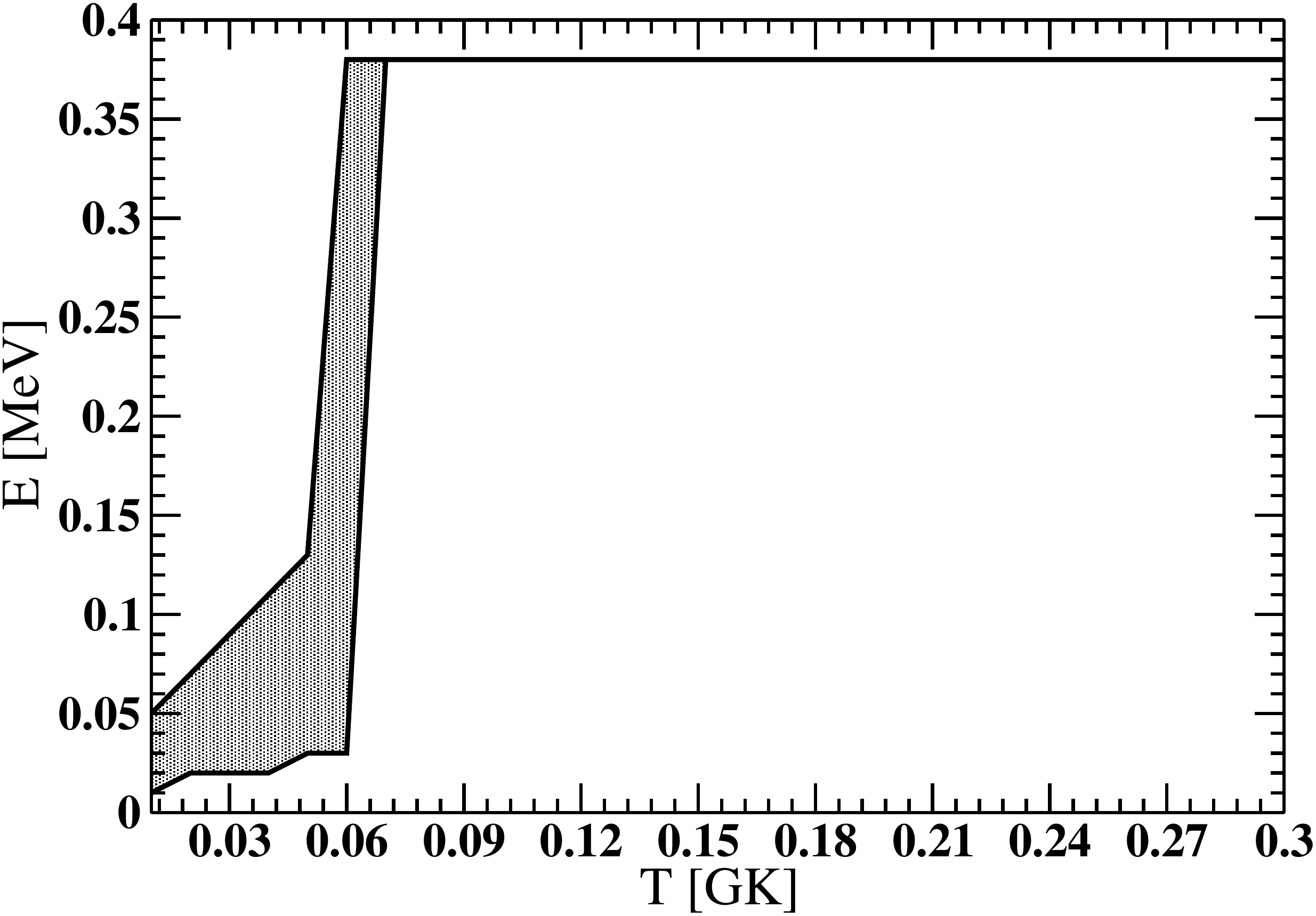}
\caption{Range of energies relevant for the triple-alpha reaction rate at a given temperature.}
\label{f10}
\end{figure}  
With the $2^+_1$ bound state and the $0^+$ continuum states of $^{12}$C obtained above, we are able to compute the triple-alpha reaction rate at different temperatures. Our method enables us to treat the resonant and non-resonant processes on the same footing. We present our new triple-alpha rate in Table~\ref{t2} for temperatures ranging from $0.01$ GK to $1.0$ GK, beyond which we would need to consider other high lying resonances. \\
\begin{table}[h!]
\caption{Our triple-alpha reaction rate (in $cm^6s^{-1}mol^{-2}$) after being normalized to NACRE for $T\ge 0.5$ GK.}
\label{t2}
\begin{ruledtabular}
\begin{tabular}{cccc}
$T$ (GK)  &$\langle R_{\alpha\alpha\alpha}\rangle$ &$T$ (GK)  &$\langle R_{\alpha\alpha\alpha}\rangle$ \\ \hline
0.010 & 8.47x10$^{-53}$ & 0.15 & 1.53x10$^{-18}$ \\
0.015& 2.11x10$^{-47}$ & 0.20  & 9.90x10$^{-16}$ \\
0.020 & 2.86x10$^{-44}$ & 0.25 & 4.13x10$^{-14}$ \\
0.025& 4.44x10$^{-42}$ & 0.30  & 4.50x10$^{-13}$ \\
0.030 & 2.08x10$^{-40}$ & 0.35 & 2.31x10$^{-12}$ \\
0.04& 5.72x10$^{-38}$ & 0.4  & 7.44x10$^{-12}$ \\
0.05 & 3.11x10$^{-36}$ & 0.5  & 3.44x10$^{-11}$ \\
0.06 & 6.79x10$^{-35}$ & 0.6  & 8.63x10$^{-11}$ \\
0.07 & 4.18x10$^{-32}$ & 0.7  & 1.55x10$^{-10}$ \\
0.08 & 7.12x10$^{-29}$ & 0.8  & 2.28x10$^{-10}$ \\
0.09 & 2.26x10$^{-26}$ & 0.9  & 2.95x10$^{-10}$	 \\
0.10  & 2.19x10$^{-24}$ & 1.0  & 3.51x10$^{-10}$ \\
\end{tabular}
\end{ruledtabular}
\end{table}
The triple-alpha reaction proceeds primarily through either the resonant or the non-resonant path depending on the temperature of the stellar environment. In order to have a full description of the triple-alpha reaction mechanism, we estimate the energy range relevant for the reaction rate at a given temperature. These ranges of energies contribute significantly to the integral of Eq.~(\ref{rateavg-eq}). As shown in Fig.~\ref{f10}, the resonant energy at $0.38$ MeV completely dominates the integration for $T\ge 0.07$ GK. For $T=0.06-0.07$ GK, there is competition between the resonant and non-resonant processes. This marks the transition region between the two processes. The non-resonant capture mechanism dominates for $T<0.06$ GK.

\section{Discussion}
\subsection{Comparison with other methods}
\begin{figure}
\center
\includegraphics[width=0.45\textwidth]{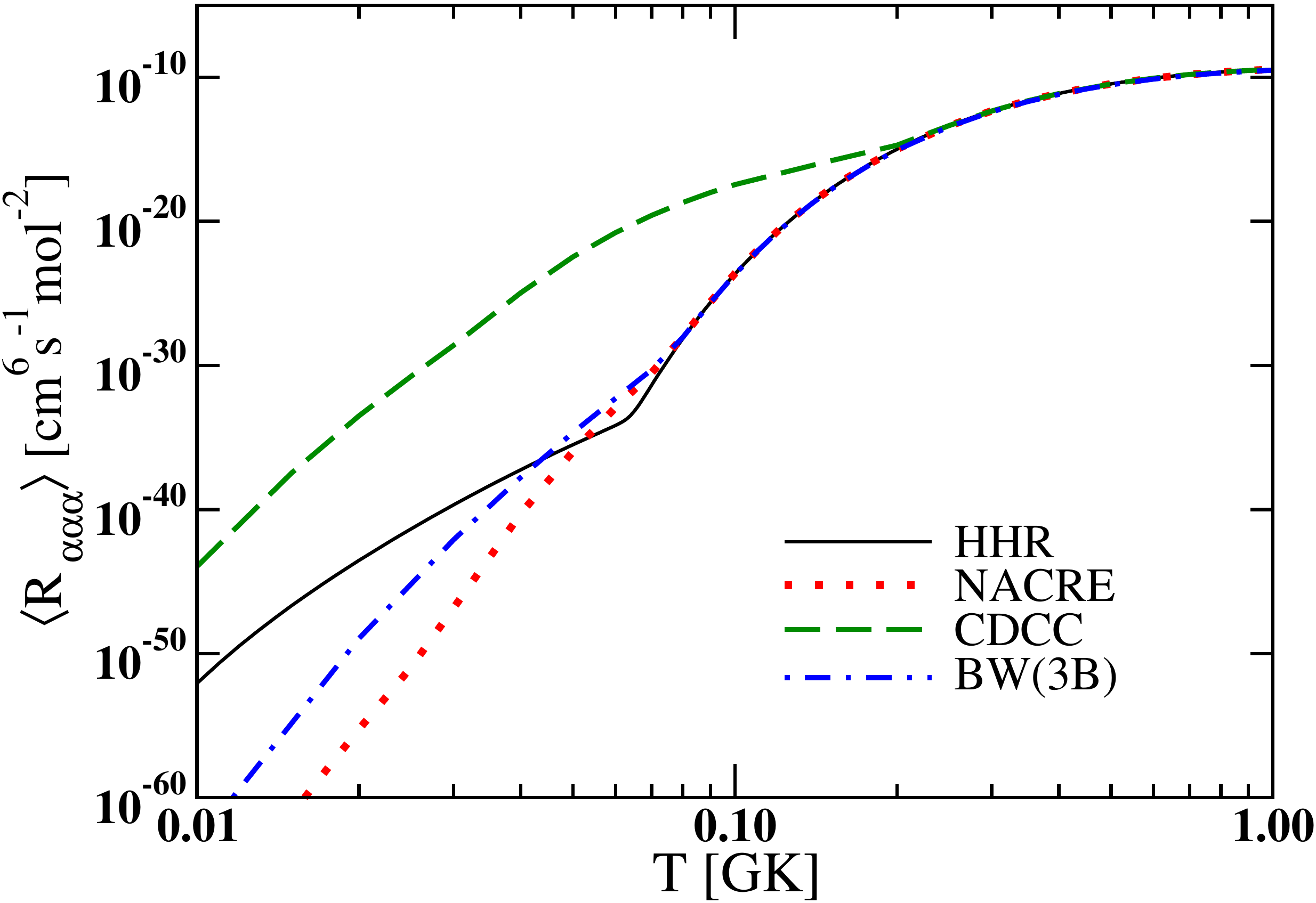}
\caption{(Color online) Different evaluations of the triple-alpha reaction rate: comparing the Hyperspherical Harmonic R-matrix method (solid) with NACRE (dotted), CDCC (dashed) and the three-body Breit Wigner (dot-dashed).}
\label{f12}
\end{figure}

Fig.~\ref{f12} presents our results (solid line) in comparison with other studies: NACRE, CDCC and BW(3B). The new rate resulting from the HHR method (after being rescaled as discussed in Sec.~\ref{conv}) agrees with NACRE (dotted line) for temperatures above $0.07$ GK. Although the reaction rate is slightly reduced below $0.07$ GK, it is significantly enhanced for $T < 0.06$ GK. We also obtained a very different temperature dependence in this region \cite{nguyen12}. The results assuming an extrapolation of a three-body Breit-Wigner cross section to low energies BW(3B) (dot-dashed line) \cite{garrido11} have a similar behavior but the reaction rate increases to a lesser extent. The CDCC results of \cite{ogata09} (long-dashed) largely enhance the triple-alpha rate for temperatures as large as $0.1$ GK when comparing with NACRE. This effect is much stronger than what is seen in our studies. The fact that the HHR result agrees with NACRE for $T>0.07$ GK leads to a negligible change in the evolution of stars around one solar mass \cite{nguyen12}. This cannot be reproduced when using the CDCC rates.\\
\indent There exists a kink in the HHR curve around $T \approx 0.06$ GK as seen in Fig.~\ref{f12}. This marks the transition between the resonant and the non-resonant processes for which the temperature dependences are different.  Above $T\approx0.06$ GK the sequential (resonant) process dominates, while below there is mostly non-sequential (direct) capture. This agrees with the finding in Fig.~\ref{f10}. The calculations in \cite{garrido11} exhibits this same feature.\\
\indent It was suggested in \cite{descouvemont10b} that the enhancement seen in the rate obtained with CDCC \cite{ogata09} is due to additional resonances in the spectrum. We have repeated the HHR calculations using a scattering wavefunction obtained from the interactions quoted in \cite{ogata09}. This interaction holds only one $0^+$ bound state at $1.686$ MeV and the converged $dB(E2)/dE$ results show no sign of additional resonances up to 1 MeV. We also repeat the HHR calculation in a truncated the model space as indicated in \cite{ogata09} (s-waves only). With such a truncated model space, the bound state becomes unbound  and
a resonance appears close to threshold (around 0.2 MeV). The resulting rate increases by many orders of magnitude.  Keeping in mind that the CDCC and the HHR methods are based on very different expansions, and the direct comparisons of intermediate steps in the calculation is not possible, our results do suggest that at least in part, the reason for the large increase in the rate seen in \cite{ogata09} is  due to model space truncation.

\subsection{Long-range Coulomb effects}
\begin{figure}
\center
\includegraphics[width=0.45\textwidth]{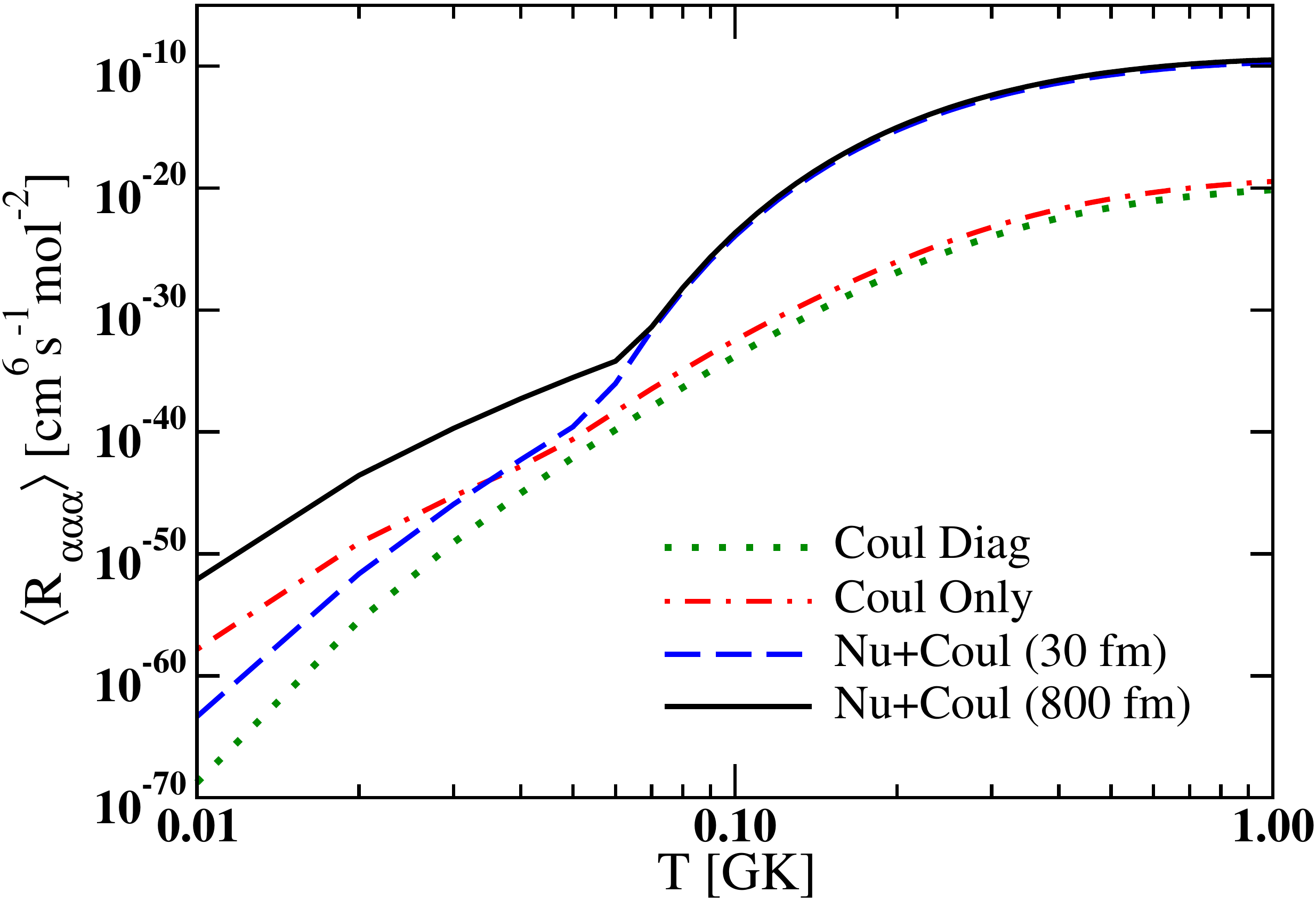}
\caption{(Color online) The long-range Coulomb effects are shown in four different calculations: only diagonal Coulomb couplings (dotted), only Coulomb couplings (dot-dashed), both nuclear and Coulomb interactions with off-diagonal Coulomb couplings up to 30 fm (dashed) and a fully converged calculation with off-diagonal Coulomb couplings up to 800 fm (solid).}
\label{f11}
\end{figure}

Since we obtain a large enhancement for the triple-alpha reaction rate at low temperatures, it is important to isolate the source of this effect. We therefore perform calculations of the $^{12}$C($0^+$) continuum states for cases in which only the diagonal Coulomb couplings $V^C_{\gamma\gamma}$ (dotted) and the full Coulomb couplings $V^C_{\gamma\gamma'}$ (dot-dashed) are included. The results are then compared to calculations including both nuclear and Coulomb interactions $ (V^C+V^N)_{\gamma\gamma'}$ with the off-diagonal Coulomb couplings up to 30 fm (dashed) and 800 fm (solid). The reaction rate for each case is then constructed by using those scattering wavefunctions and fixing the $^{12}$C($2^+_1$) bound state. Fig.~\ref{f11} shows the results of these calculations. When only the diagonal Coulomb couplings are present, we are able to obtain an analytic solution of Eq.~(\ref{hh-eq}) (Sec.~\ref{HHform}). Our numerical calculation for this case agrees well with the analytic solution allowing us to test our implementation. The inclusion of the off-diagonal Coulomb couplings (dot-dashed curve) significantly increases the reaction rate at low temperatures. When both nuclear and Coulomb couplings are fully included in our calculation, we observe an increase in the reaction rate at high temperatures due to the resonant contribution. Comparision between the solid and the dashed curves confirms that off-diagonal Coulomb couplings drive the increase in the triple-alpha reaction rate at low temperatures. Fig.~\ref{f11} indicates that the effect of off-diagonal long-range couplings are relatively small at high temperatures but very important in the low temperature regime. About $10$ orders of magnitude enhancement in the rate is found at $T=0.01$ GK due to these effects, demonstrating the importance of including Coulomb correctly.\\
\subsection{Reaction dynamics}
\begin{figure}
\center
\includegraphics[width=0.45\textwidth]{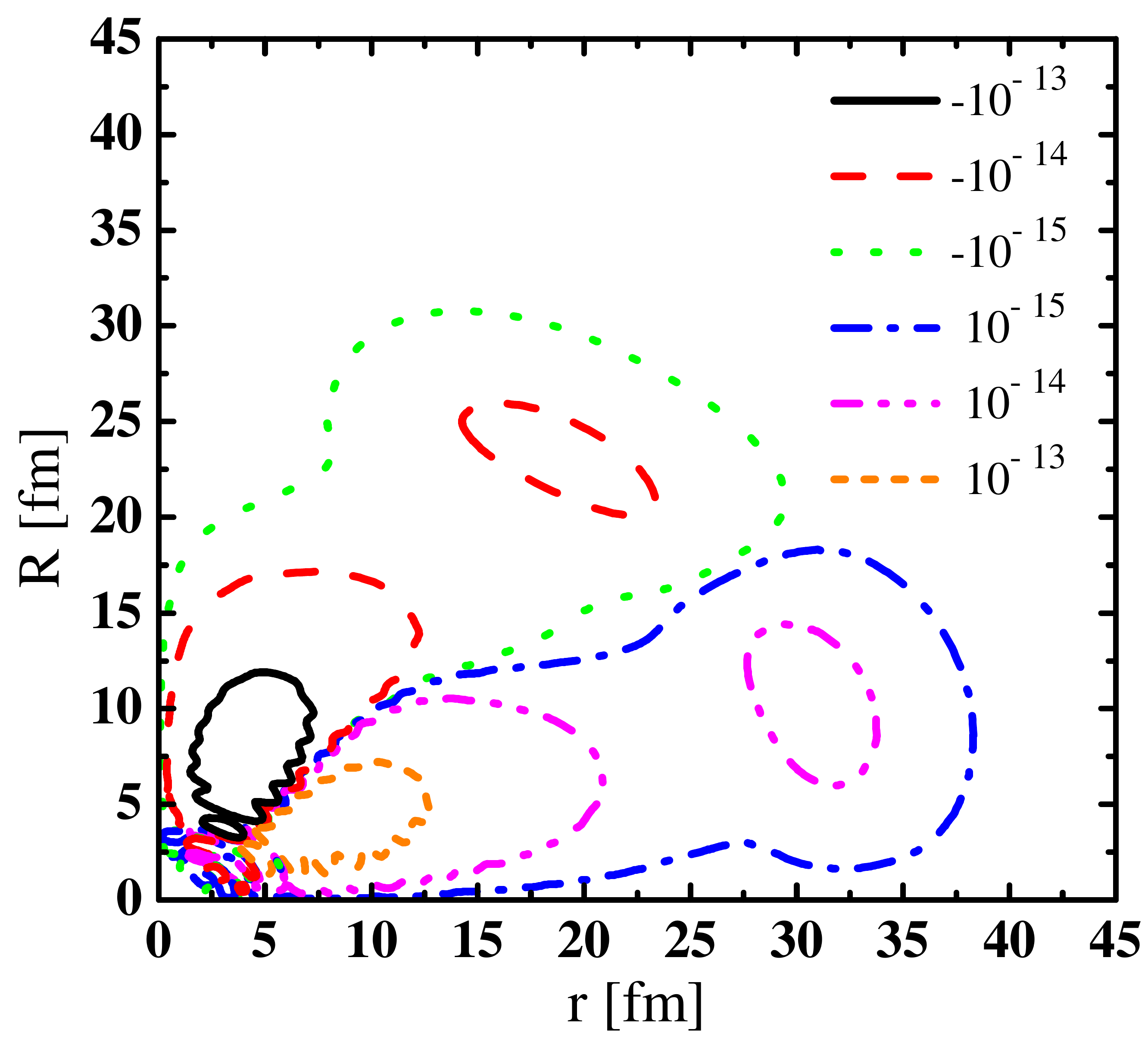}
\caption{(Color online) Spatial distribution of function $f_{\gamma^i}(r,R)$ for $E=0.05$ MeV and $\gamma^i=\{0, 0, 0\}$.}
\label{f14}
\end{figure}

\begin{figure}
\center
\includegraphics[width=0.45\textwidth]{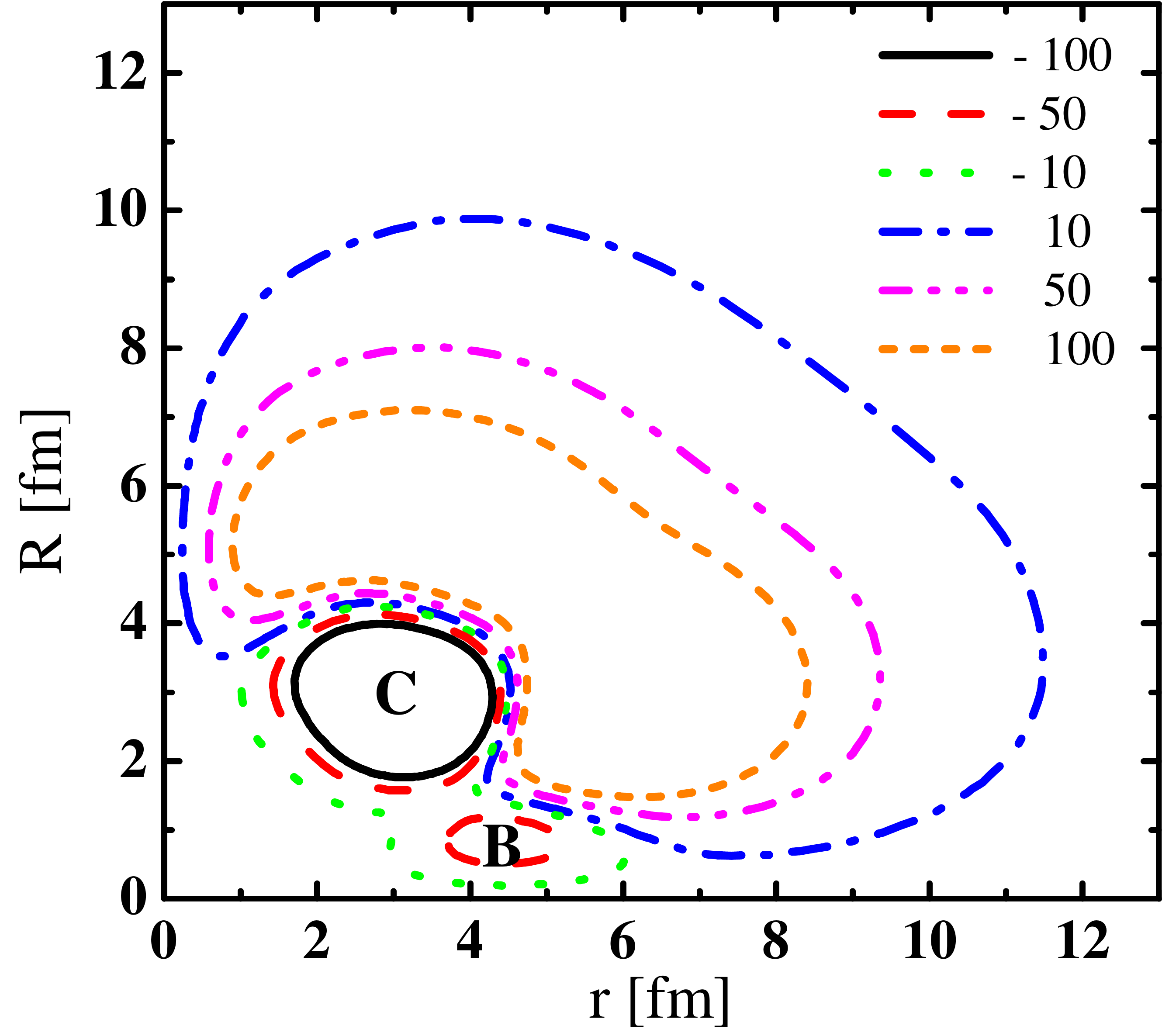}
\caption{(Color online) Spatial distribution of function $f_{\gamma^i}(r,R)$ for $E=0.38$ MeV and $\gamma^i=\{0, 0, 0\}$.}
\label{f15}
\end{figure}

\begin{figure}
\center
\includegraphics[width=0.45\textwidth]{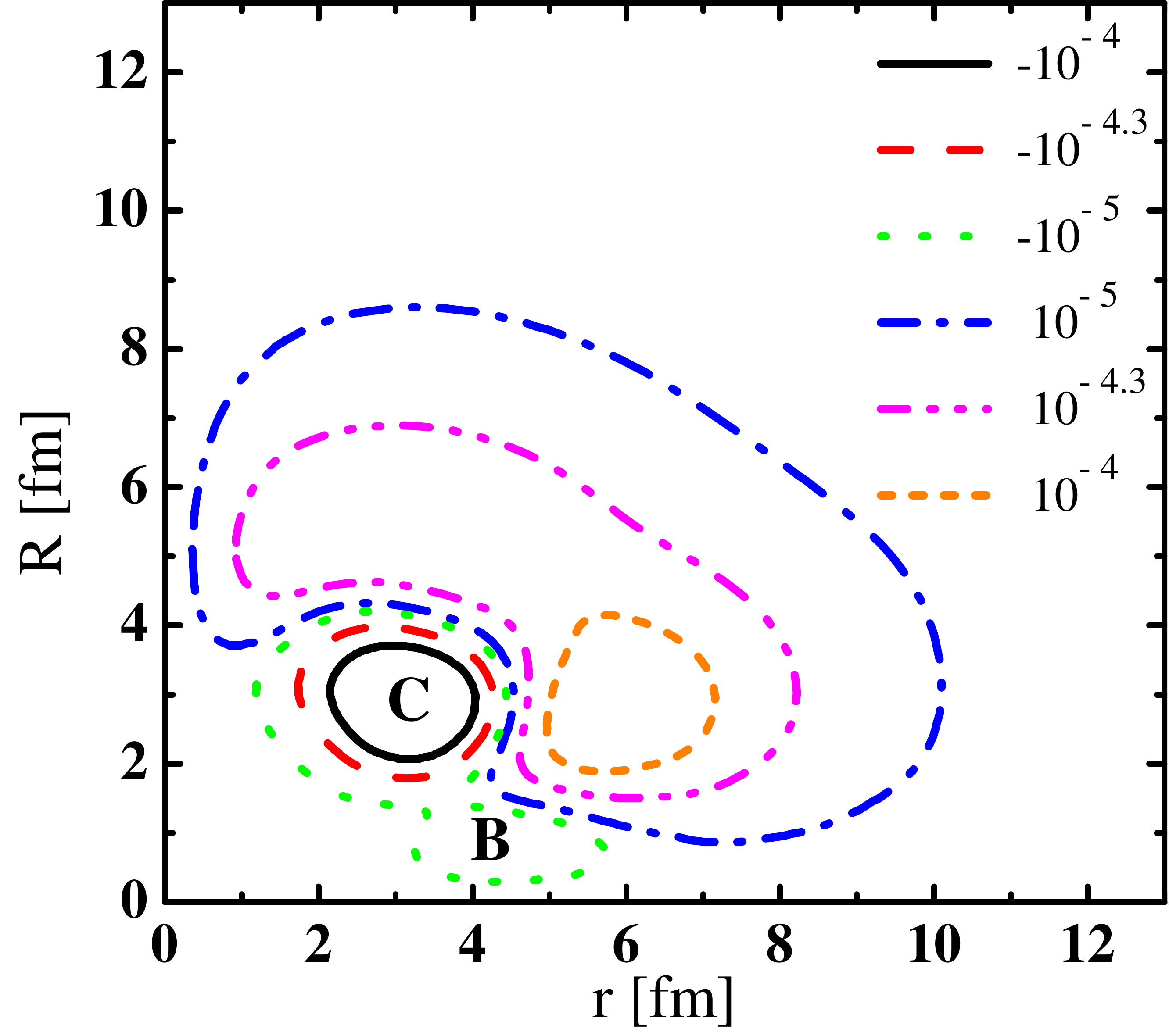}
\caption{(Color online) Spatial distribution of function $f_{\gamma^i}(r,R)$ for $E=0.5$ MeV $\gamma^i=\{0, 0, 0\}$.}
\label{f16}
\end{figure}

In order to understand the mechanism for $^{12}$C production in more detail, we rewrite Eq.~(\ref{be2e-eq}) as
\begin{equation}
\frac{\mathrm{d}B(E2)}{\mathrm{d}E}\sim\left|\sum_{\gamma^i}\int f_{\gamma^i}(r,R)\mathrm{d}r\mathrm{d}R\right|^2\, ,
\end{equation}
where $r$ is the radius between two alpha particles and $R$ is the distance from their center of mass to the third alpha particle. $\gamma^i = \{K^i, l_x^i, l_y^i\}$ indicates an incoming channel in the hyperspherical wave expansion for a scattering state. The spatial distribution of function $f_{\gamma^i}(r,R)$ at different three-body kinetic energies $E$ contains information about the dynamics of the triple-alpha reaction. Fig.~\ref{f14},~\ref{f15}, and~\ref{f16} illustrate the spatial distributions of $f_{\gamma^i}(r,R)$ at a very low energy $E=0.05$ MeV, at the resonant energy $E=0.38$ MeV, and at an energy well above the resonance $E=0.5$ MeV, respectively. We just present here the distribution functions corresponding to the first incoming channel ($K^i=0$, $l_x^i=0$, and $l_y^i=0$) which is the dominant contribution to the quadrupole strength function $\mathrm{d}B(E2)/\mathrm{d}E$. Other channels exhibit the same trends.\\
\indent At low energy $E=0.05$ MeV, the spatial distribution of function $f_{\gamma^i}(r,R)$ shown in Fig.~\ref{f14} has a different symmetry in comparison with the higher energies (Fig.~\ref{f15},~\ref{f16}). We observe comparable contributions coming from two different triple-alpha configurations: the prolate triangle and the oblate triangle as shown in Fig.~\ref{config}a and Fig.~\ref{config}b, respectively. There is a large cancellation between these two contributions resulting in a small value of $\mathrm{d}B(E2)/\mathrm{d}E$ at low energies (for example, the two contributions are $\sim 10^{-12}$ but their sum is $\sim 10^{-16}$). Nevertheless these cancellations are well within the numerical accuracy of our computations. \\
\indent Fig.~\ref{f15} and Fig.~\ref{f16} illustrate the spatial distribution of function $f_{\gamma^i}(r,R)$ at the resonant energy $E=0.38$ MeV and higher $E=0.5$ MeV. These two cases share the same symmetry. The dominant contribution to the quadrupole strength function $\mathrm{d}B(E2)/\mathrm{d}E$ comes from the region within the smallest contour in Fig.~\ref{f15} and Fig.~\ref{f16} that contains both maxima {\bf B} and {\bf C}. The maximum {\bf C} in the spatial distribution of function $f_{\gamma^i}(r,R)$ is caused by the triple-alpha equilateral triangle configuration in both the $2^+_1$ bound state and the Hoyle resonant state (see the density distribution for each state in Fig.~\ref{2bsdens} and Fig.~\ref{f5}). The contribution to the maximum {\bf B} mostly comes from the three-alpha oblate configuration which only appears in the Hoyle resonant state. Even though the three-alpha prolate configuration dominates the Hoyle state's structure, it does not contribute significantly to the strength function because the $2^+_1$ bound state wavefunction is zero in that region.
\section{Conclusion}
We have successfully produced the triple-alpha reaction rate in the low temperature regime $T<0.1$ GK where many numerical attempts have failed before. In this paper, the triple-alpha is modeled as a three-body Borromean system. We employ the hyperspherical harmonics (HH) method to tackle this problem. In the low temperature region, the triple-alpha proceeds through a quadrupole transition from the $0^+$ continuum to the $2^+_1$ bound state in $^{12}$C. The $2^+_1$ bound state is obtained by solving a set of coupled channels equations in hyper-radius coordinates for negative energy and the condition that the wavefunction goes to zero at large distances. Naturally the same approach cannot be applied to the $0^+$ continuum states. We combine the R-matrix expansion and R-matrix propagation method in the hyperspherical harmonics basis to obtain the solutions for the continumm states. We also implement a technique of screening the off-diagonal Coulomb couplings to ensure numerical stability at very low energies.\\
\indent The HHR method enables us to treat the resonant and the non-resonant process on the same footing. We construct the triple-alpha reaction rate from the quadrupole transition from the $0^+$ continuum states to the $2^+_1$ bound state of $^{12}$C. A thorough convergence study is performed. Our reaction rate converges well with screening radius and the size of our model space. We estimate an overall uncertainty of a factor of 2 for our triple-alpha rate. The new rate agrees with NACRE above $0.07$ GK. However we find a large enhancement of the rate at temperatures below $0.06$ GK which marks the transition between the resonant and the non-resonant process. Although we obtain a significant increase of the rate at low temperatures, it does not drastically change the evolution of low-mass stars \cite{nguyen12} as was the case for the CDCC rates \cite{dotter09}. Our rate preserves the red giant phase and is therefore consistent with observations \cite{nguyen12}. We expect our new rate to have larger impact on some astrophysical scenarios that burn helium at lower temperatures, such as helium-accreting white dwarfs and neutron stars with small accretion rate. The astrophysics implications are not presented in this paper. A separate publication with more details of astrophysical impacts will appear in the near future.\\
\indent We also explore the importance of long-range effects in our problem by performing four calculations in which only the diagonal Coulomb couplings, the full Coulomb couplings, and both nuclear and Coulomb interactions with the off-diagonal Coulomb couplings up to 30 fm and 800 fm are included. The results emphasize the necessity to include Coulomb effects correctly, especially at low temperatures.\\
\indent The presence of very narrow two-body and three-body resonances in addition to the strong, long-range Coulomb interaction make the triple-alpha problem very challenging. The HHR framework which is the combination of various methods is a new approach to overcome the well known difficulty of the three charged particle system. This method allows us to approach the very low energy regime where measurements are impossible without using extrapolation, resulting in significant impacts in astrophysics. In addition, it opens new opportunities in addressing three-body low-energy reactions in others fields for cases where all particles have charge.

\begin{acknowledgments}
We thank Edward Brown, Richard Cyburt, Ron Johnson, Akram Mukhamedzhanov and Chuck Horowitz for useful discussions during this project. This work was supported by the National Science Foundation grant PHY-0800026 and the Department of Energy under contracts DE-FG52-08NA28552 and DE-SC0004087. This work was performed in part under the auspices of the U.S. Department of Energy by Lawrence Livermore National Laboratory under Contract DE-AC52-07NA27344.

\end{acknowledgments}
\appendix
\section{Quadrupole transition function}
In this appendix, we present a sumary of our derivation of the three-body quadrupole transition function in hyperspherical coordinates. Our problem involves only spin-zero particles, notation $L$ is therefore used to refer to the total angular momentum and the total spin of the system. A transition from the initial continuum state of total spin $L$ and momentum $\kappa$ to the final bound state of spin $L'$ is characterized by the strength function:
\begin{eqnarray}
&&\frac{\mathrm{d}B(E2,L\rightarrow L')}{\mathrm{d}E}=\frac{E^2}{2}\:\left(\frac{2m}{\hbar^2}\right)^3\nonumber\\
&&\times\int\sum_{mM'M}\left|\langle L'M'\left |E_{2m}\right|LM; \kappa\rangle\right|^2 \mathrm{d}\Omega^{\kappa}_5\, .\label{be2e-eq}
\end{eqnarray}
The integration in Eq.~(\ref{be2e-eq}) is taken over all angles in momentum space $\mathrm{d}\Omega^{\kappa}_5=\mathrm{sin}\theta_{\kappa}\:\mathrm{cos}\theta_{\kappa}\:\mathrm{d}\theta_{\kappa}\:\mathrm{d}\Omega_{k_x}\:\mathrm{d}\Omega_{k_y}$. The final bound state wavefunction $\Psi^{L'M'}$ is expanded in HH coordinates as:
\begin{equation}
\Psi^{L'M'}=\rho^{-\frac{5}{2}}\sum_{K'l'_xl'_y}\chi_{K'l'_xl'_y}(\rho)\:\varphi_{K'}^{l'_xl'_y}(\theta)\:[Y_{l'_x}\otimes Y_{l'_y}]_{L'M'} \, ,\label{bsexp-eq}
\end{equation}
while the continuum state wavefunction $\Psi^{LM}$ has the following formula:
\begin{eqnarray}
\Psi^{LM}=\frac{1}{(\kappa\rho)^{5/2}}\sum_{\substack{ K^il^i_xl^i_y\\ Kl_xl_y}}&&\chi^{K^il^i_xl^i_y}_{Kl_xl_y}(\kappa\rho)\phi_{K}^{l_xl_y}(\theta)[Y_{l_x}\otimes Y_{l_y}]_{LM}\nonumber\\
&&\times\phi_{K^i}^{l^i_xl^i_y}(\theta_{\kappa})[Y_{l^i_x}\otimes Y_{l^i_y}]^{\kappa}_{LM}\, . \label{contexp-eq}
\end{eqnarray} 
In Eq.~(\ref{contexp-eq}), $\kappa$ is the hyper-radial momentum and relates to the three-body energy $E$ by a relationship $\kappa=\sqrt{2mE/\hbar^2}$; $K^il^i_xl^i_y$ ($Kl_xl_y$) represents the incoming (outgoing) channel. We first expand the squared modulus sum in Eq.~(\ref{be2e-eq}) using Eq.~(\ref{bsexp-eq}) and Eq.~(\ref{contexp-eq}). The integration in momentum space $\mathrm{d}\Omega^{\kappa}_5$ is then simplified using the following properties: 
\begin{align}
&\int \mathrm{d}\Omega_{k_x}Y^*_{l^i_x m^i_x}(\Omega_{k_x})Y_{l^i_{1x} m^i_{1x}}(\Omega_{k_x})=\delta_{l^i_x l^i_{1x}}\delta_{m^i_x m^i_{1x}}\, ,\label{y1-eq}\\
&\int \mathrm{d}\Omega_{k_y}Y^*_{l^i_x m^i_x}(\Omega_{k_y})Y_{l^i_{1y} m^i_{1y}} (\Omega_{k_y})=\delta_{l^i_y l^i_{1y}}\delta_{m^i_y m^i_{1y}}\, ,\label{y2-eq}\\
&\int \mathrm{sin}^2\theta_{\kappa} \mathrm{cos}^2\theta_{\kappa} \mathrm{d}\theta_{\kappa}(\phi_{K^i}^{l^i_xl^i_y}(\theta_{\kappa}))^*\phi_{K^i_1}^{l^i_xl^i_y}(\theta_{\kappa})=\delta_{K^i,K^i_1}\, .\label{phi-eq}
\end{align}
We arrive at a simpler expression for Eq.~(\ref{be2e-eq}) which no longer contains the momentum space dependence
\begin{align}
&\frac{\mathrm{d}B(E2,L\rightarrow L')}{\mathrm{d}E}=\frac{E^2}{2}\left(\frac{2m}{\hbar^2}\right)^3\frac{\hat{L'}^2}{\hat{L}^2}\sum_{K^il^i_xl^i_y}\nonumber\\
&\left|\sum_{Kl_xl_y}\sum_{K'l_x'l_y'}\langle K'l_x'l_y'L'\left |\left|E_{2m}\right|\right|Kl_xl_yK^il^i_xl^i_yL\rangle_s\right|^2, \label{dbe2-eq}
\end{align}
where the subscript s denotes the radial-space part in the wavefunction expansion and $\hat{L}^2=2L+1$. \\
It is straight forward to obtain the expression for the quadrupole operator $E_{2m}$ for three alpha particles in HH coordinates:
\begin{equation}
E_{2m}=\frac{eZ}{A}\:\left[(\rho\: \mathrm{sin}\theta)^2\:Y_{2m}(\hat{\bf x})+(\rho\: \mathrm{cos}\theta)^2\:Y_{2m}(\hat{\bf y})\right]\, .
\label{e2op-eq}
\end{equation}
Inserting Eq.~(\ref{e2op-eq}) into Eq.~(\ref{dbe2-eq}) we have:
\begin{eqnarray}
&&\langle K'l_x'l_y'L'\left |\left|E_{2m}\right|\right|Kl_xl_yK^il^i_xl^i_yL\rangle_s=\label{mtbe2b-eq}\\
&&\frac{eZ}{A}\:\langle K'l_x'l_y'L'\left |\left|(\rho\: \mathrm{sin}\theta)^2Y_{2}(\hat{x})\right|\right|Kl_xl_yK^il^i_xl^i_yL\rangle_s\nonumber\\
&&+\frac{eZ}{A}\:\langle K'l_x'l_y'L'\left |\left|(\rho\: \mathrm{cos}\theta)^2Y_{2}(\hat{y})\right|\right|Kl_xl_yK^il^i_xl^i_yL\rangle_s\, .\nonumber
\end{eqnarray} 
We denote $M.E.1$ and $M.E.2$ as the first and second term on the $r.h.s$ of Eq.~(\ref{mtbe2b-eq}). $M.E.1$ can be factorized into two terms of which one contains the hyperspherical variable dependence and the other is angular momentum dependent:
\begin{eqnarray}
M.E.1&&=\frac{eZ}{A}\:\langle K'l_x'l_y'\left|(\rho\:\mathrm{sin}\theta)^2\right|Kl_xl_yK^il^i_xl^i_y\rangle\nonumber\\
&&\times\langle l_x'l_y'L'\left|\left|Y_{2}(\hat{x})\right|\right|l_xl_yL\rangle.\label{quadtran1-eq}
\end{eqnarray}
The first bra-ket term on the $r.h.s$ of Eq.~(\ref{quadtran1-eq}) is calculated by taking an integral over the hyper-radial and hyper-angular parts of the wavefunctions:
\begin{equation}
\langle K'l_x'l_y'\left|(\rho\:\mathrm{sin}\theta)^2\right|Kl_xl_yK^il^i_xl^i_y\rangle=\left(\frac{2mE}{\hbar^2}\right)^{-5/4}I_{\rho}I_{1,\theta}\, ,\label{rhotheta1-eq}
\end{equation}
where we define:
\begin{align}
I_{\rho}&=\int \mathrm{d}\rho \:\chi_{K'l_x'l_y'}^*(\rho)\:\rho^2\:\chi_{Kl_xl_y}^{K^il^i_xl^i_y}(\rho)\, ,\label{irho-eq}\\
I_{1,\theta}&=\int \mathrm{sin}^2\theta \:\mathrm{cos}^2\theta\: \mathrm{d}\theta\: (\phi_{K'}^{l_x'l_y'}(\theta))^*\mathrm{sin}^2\theta\: \phi_K^{l_xl_y}(\theta)\label{itheta1-eq}\, .
\end{align}
The second bra-ket term on the $r.h.s$ of Eq.~(\ref{quadtran1-eq}) can be explicitly calculated using angular momentum algebra \cite{brink94} as:
\begin{eqnarray}
&&\langle l_x'l_y'L'\left|\left|Y_{2}(\hat{x})\right|\right|l_xl_yL\rangle=\delta(l_y,l_y')\sqrt{\frac{5}{4\pi}}\hat{L}\hat{l_x'}\hat{l_x}\nonumber\\
&&\times(-1)^{l_y'+L}\left\{\begin{array}{ccc} L'& L& 2\\l_x& l_x'& l_y'\end{array}\right\}\left(\begin{array}{ccc} l_x'& 2& l_x\\0& 0& 0\end{array}\right).\quad\label{SH1}
\end{eqnarray}
Performing a similar calculation for the second term on the $r.h.s$ of of Eq.~(\ref{mtbe2b-eq}) ($M.E.2$), we obtain the final result for a three-body quadrupole transition strength function:
\begin{eqnarray}
&&\frac{\mathrm{d}B(E2,L\rightarrow L')}{\mathrm{d}E}=\sqrt{\frac{m}{2\hbar^2}}\:\frac{1}{\sqrt{E}}\:\hat{L'}^2\left(\frac{eZ}{A}\right)^2\nonumber\\
&&\times\sum_{K^il^i_xl^i_y}\left|\sum_{Kl_xl_y}\sum_{K'l_x'l_y'}I_{\rho}\left(I_{1,\theta}A_1+I_{2,\theta}A_2\right )\right |^2, \label{dbe2f-eq}
\end{eqnarray}
where $I_{\rho}$ and $I_{1,\theta}$ are defined in Eq.~(\ref{irho-eq}) and Eq.~(\ref{itheta1-eq}) respectively. Other quantities are given as:
\begin{eqnarray}
&&I_{2,\theta}=\int \mathrm{sin}^2\theta \: \mathrm{cos}^2\theta\:\mathrm{d}\theta \:(\phi_{K'}^{l_x'l_y'}(\theta))^*\:\mathrm{cos}^2\theta \:\phi_K^{l_xl_y}(\theta),\nonumber\\
&&A_1=\delta(l_y,l_y')\sqrt{\frac{5}{4\pi}}\hat{l_x}\hat{l_x'}(-1)^{l_y'+L} \nonumber\\
&&\quad\quad\times\left\{\begin{array}{ccc} L'& L& 2\\l_x& l_x'& l_y'\end{array}\right\}\left(\begin{array}{ccc} l_x'& 2& l_x\\0& 0& 0\end{array}\right),\nonumber\\
&&A_2=\delta(l_x,l_x')\sqrt{\frac{5}{4\pi}}\hat{l_y}\hat{l_y'}(-1)^{l_x'+L'+l_y'+l_y} \nonumber\\
&&\quad\quad\times\left\{\begin{array}{ccc} L'& L& 2\\l_y& l_y'& l_x'\end{array}\right\}\left(\begin{array}{ccc} l_y'& 2& l_y\\0& 0& 0\end{array}\right).\nonumber
\end{eqnarray}



\end{document}